\documentclass{aa}  

\usepackage{color}
\usepackage{graphicx}
\usepackage{txfonts}
\usepackage{hyperref}

\usepackage[utf8]{inputenc}
\usepackage{amsmath}
\usepackage{amssymb}
\usepackage{physics}
\usepackage[mathlines]{lineno}

\allowdisplaybreaks[4]
\begin{document} 

\title{Modelling gyrosynchrotron emission from coronal energetic electrons in a CME flux rope}

   \author{E. Husidic
          \inst{1,2}
          \and 
          N. Wijsen
          \inst{1} 
          \and 
          I.~C. Jebaraj
           \inst{2} 
           \and 
          A. Vourlidas
           \inst{3} 
           \and
           L. Linan
           \inst{1}
           \and
          R. Vainio
           \inst{2}
           \and
           S. Poedts
           \inst{1,4}
          }

   \institute{Centre for mathematical Plasma Astrophysics, Department of Mathematics, KU Leuven, Celestijnenlaan 200B, 3001 Leuven, Belgium\\
              \email{edin.husidic@kuleuven.be}
         \and
             Department of Physics and Astronomy, University of Turku, 20014 Turku, Finland
        \and
             The Johns Hopkins University Applied Physics Laboratory,
             11001 Johns Hopkins Rd, 20723 Laurel, USA
         \and
             Institute of Physics, University of Maria Curie-Sk{\l}odowska, Pl.\ Marii Curie-Sk{\l}odowskiej 1, 20-031 Lublin, Poland
             }

   \date{Received: 16 May 2025; Accepted: 21 July 2025}

  \abstract
   {Solar flares and coronal mass ejections (CMEs) can accelerate electrons, causing bursts such as type~IV emissions in the solar radio continuum. Although radio spectroscopy is a powerful diagnostic tool for the corona, the origin and mechanisms of type~IV bursts remain uncertain. In situ measurements can occasionally shed some light on these mechanisms, but they are limited in space and time. Sophisticated numerical modelling offers the best approach to improve our understanding of the physical processes involved.}
   {This research examines type~IV radio bursts, exploring the effects of various electron distribution properties and CMEs on their generation and characteristics. To transcend idealised assumptions, we employ realistic, anisotropic electron distributions --- obtained from particle transport simulations within complex magnetohydrodynamic (MHD) environments --- as input for radio emission models.}
   {We use the three-dimensional coronal MHD model COCONUT to generate coronal background configurations, including a CME modelled as an unstable modified Titov--D\'{e}moulin magnetic flux rope (MFR). These MHD simulations are used by the PARADISE particle transport code, which injects energetic electrons into the MFR and tracks their evolution. Finally, we feed the electron distributions and solar wind parameters into the Ultimate Fast Gyrosynchrotron Codes to compute radio emission along lines of sight.}
    {Electrons injected close to the flux rope's central axis remain largely confined, producing a gyrosynchrotron emission spectrum resembling observed type~IV characteristics. Varying observer positions, CME properties, and spectral indices of the electron energy distributions modify the intensities and durations of the observed bursts. The strongest gyrosynchrotron emission is observed to originate from the CME flanks.}
   {Our results indicate that gyrosynchrotron emission is the major component in type~IV spectra, although additional contributors cannot be ruled out.
   }
   
   \keywords{Sun: corona -- Sun: particle emission -- Sun: coronal mass ejections (CMEs) -- Sun: radio radiation}

\titlerunning{Modelling electron gyrosynchrotron emission}
\authorrunning{Husidic et al.}
\maketitle
\section{Introduction}\label{sec:introduction}

The solar corona is a highly dynamic environment, characterised by spatio-temporal variations that manifest across its broadband electromagnetic spectrum, extending from radio wavelengths to gamma rays. Fundamental plasma processes in the corona give rise to diverse radiative phenomena, broadly categorised by how individual electrons contribute to the observed radiation \citep{Kundu-1965}. A common classification distinguishes between coherent and incoherent emissions \citep{Kaplan-Tsytovich-1969}. In coherent emission, many thermal or nonthermal electrons act collectively, producing phase-correlated waves. This collective behaviour can lead to highly amplified and organised bursts of radiation \citep{Zheleznyakov-1970, Melrose-1980, Melrose-2017}. In contrast, incoherent emission occurs when electrons radiate independently with random phases, resulting in more gradual and less intense radiation \citep{Schwinger-1949, Born-Wolf-1959, Ginzburg-Syrovatskii-1964}. These classifications relate to the underlying emission mechanisms. Spontaneous emission, for instance, occurs when electrons radiate independently, typically resulting in incoherent radiation. Induced (or stimulated) emission, on the other hand, happens when incident radiation triggers electrons to emit additional phase-aligned radiation, often leading to coherent bursts \citep{Galeev-etal-1965,Litvak-Trakhtengerts-1971,Papadopoulos-Freund-1979}. 

Under typical coronal conditions, and more broadly in space plasmas, characteristic plasma frequencies, such as the electron plasma frequency and the electron gyrofrequency, fall within the radio regime, making radio observations a powerful diagnostic tool \citep{Nindos-2020}. Spontaneous, incoherent processes generally dominate in thermal plasmas. In contrast, spontaneous and induced processes may occur in nonthermal environments, such as during solar flares, giving rise to a rich variety of radio emission features \citep{Pick-Vilmer-2008}. This diversity of solar radio emission is categorised into four primary types, formally labelled as types I -- IV. During extreme releases of the Sun's magnetic energy, whether through solar flares or coronal mass ejections (CMEs), all types of radio emissions may be observed, providing a unique window into high-energy physics that remains inaccessible at other wavelengths \citep{Knock-Cairns-2005, Jebaraj-etal-2021, Alissandrakis-etal-2021, Klein-etal-2022,Jebaraj-etal-2023, Jebaraj-etal-2024, Wilson-etal-2025}. This makes radio emission a valuable probe of the acceleration, release, and transport of solar energetic particles (SEPs). However, only energetic electrons directly produce radio emissions that are detectable from the solar atmosphere \citep{Kouloumvakos-etal-2015}. 

\begin{figure}
    \centering
\includegraphics[width=0.9\columnwidth]{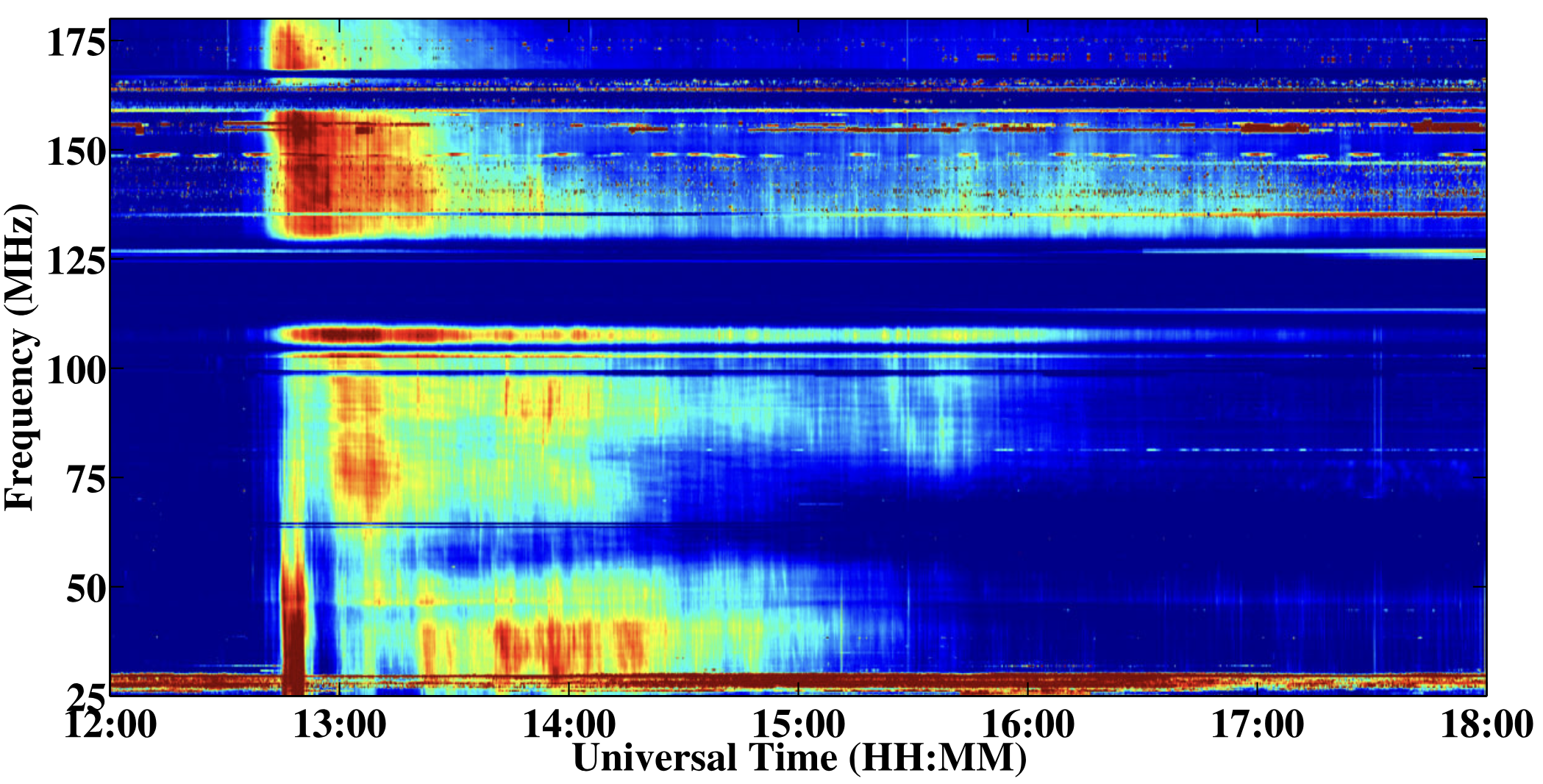}
    \caption{Dynamic radio spectrum of a type~IV radio burst detected by the Sagamore Hill Solar Radio Observatory on 24 September 2011. Adapted from \cite{Kumari-2022}.}
    \label{fig:observational_data}
\end{figure}

Solar radio emissions have been studied extensively (e.g., \citealt{Melrose-1980,Gary-Keller-2004}), but several fundamental questions remain. An enduring conundrum involves the origin of type~IV bursts, associated with energetic eruptions. The name type~IV refers to both a broadband continuum and the bursty emission that may occur within it. It is important to distinguish the two because their emission mechanisms are believed to be fundamentally different and to occur under distinct conditions. Figure~\ref{fig:observational_data} shows a dynamic radio spectrum of a type~IV burst recorded by the Sagamore Hill Radio Observatory on 24 September 2011 \citep{Kumari-2022}. The spectrum displays several features commonly observed in type~IV bursts, including a gradual drift of the peak intensity toward lower frequencies over time (above 125~MHz), stationary (in frequency) structures (between 25~and 50~MHz), and a bursty, high-intensity core surrounded by regions of weaker emission. While the depicted type~IV burst lasted around 5~hours, typical durations of type~IVs range from minutes to about an hour \citep{Robinson-1978, Gergely-1986}.

Type~IV emissions are subdivided into moving and stationary spectral types, referred to as type~IVm and type~IVs, respectively \citep[e.g.,][and references therein]{Vourlidas-2004}. Type~IVm can be synchrotron, gyrosynchrotron (GS) or plasma emission of electrons entrained within CMEs \citep[e.g.,][]{Dulk-1973}. Type~IVs radio bursts are usually attributed to plasma emission \citep{Melrose-1980}. 
Historically, spectral drift has often been interpreted as a proxy for radio source motion, with spectral drift implying spatial motion of the emission region \citep{Weiss-1963, Robinson-1978, Zlotnik-etal-2003, Chernov-2006, Nishimura-etal-2013, Bain-etal-2014, Kumari-etal-2021}, and the absence of spectral drift suggesting a stationary source \citep{Weiss-1963, Robinson-Smerd-1975, Dulk-1985, Lv-etal-2021}. However, it should be noted that imaging observations have suggested that both (spectrally defined) IVm and IVs continua can exhibit spatial drift \citep{Koval-etal-2016, Salas-Matamoros-Klein-2020, Morosan-etal-2021}. In addition, \cite{Lv-etal-2021} report frequency-dependent spatial structuring for IVs events, possibly reflecting embedded CMEs. Furthermore, \cite{Kumari-etal-2021} found in their study that while most CMEs were accompanied by type IV bursts, only a minority of these were actually of the moving type.
If such a spectral and spatial decorrelation exists in type~IV behaviour, it must be an intrinsic property of the source itself.

Nonthermal electrons accelerated during CME eruptions can emit GS radiation if they remain trapped within an erupting flux rope or plasmoid. This type of emission, sometimes referred to as a "radio CME," differs from the radiation originating at shock fronts or from remote flare loops. Relativistic electrons (with Lorentz factors \(\gamma \gtrsim 1\), where $\gamma = (1 - v^2/c^2)^{-1/2}$) spiralling along solar magnetic field lines produce GS radiation, whereas ultrarelativistic electrons (\(\gamma \gg 1\)) generate synchrotron emission. According to standard theory, GS emission occurs in magnetoactive plasma environments where dispersive effects significantly modify the radiation and must be considered \citep{Melrose-1968,Zheleznyakov-1969}. In contrast, pure synchrotron emission is typically treated as a vacuum process, appropriate when plasma densities are low or emission frequencies high enough to minimise medium-related effects \citep{Schwinger-1949,Ginzburg-1979}.

For these reasons, type IVm bursts have an important space weather aspect as they carry information of the magnetic content of the CME \citep{Vourlidas-etal-2020}. However, they have been difficult to detect and study. Coherent emissions tend to dominate the radio spectra during eruptions, hindering the detection of faint emission from the CME interior. Joint white-light and radio observations of transients have been rare until the launch of the Solar and Heliospheric Observatory (SOHO) in the mid-1990s. Soon after, \cite{Bastian-etal-2001} reported the first direct radio imaging of a white-light CME, revealing a moving radio source within the CME with a spectrum consistent with GS emission. The study obtained the spatial distribution of energetic electrons within the CME and an estimate of the CME’s internal magnetic field strength and electron population. Since then, several additional events have been detected \citep{Maia-etal-2007,Tun-Vourlidas-2013, Mondal-etal-2020, Kansabanik-etal-2024}.

A key research focus in type~IV bursts is using their emission to estimate CME magnetic fields, which are nearly impossible to assess by other means \citep{Morosan-etal-2019, Kumari-etal-2021}. Various studies have interpreted GS emission as the primary mechanism in type~IV radio bursts to infer magnetic field strengths within CMEs at specific solar radial distances \citep{Gopalswamy-Kundu-1987, Bastian-etal-2001, Maia-etal-2007, Tun-Vourlidas-2013, Bain-etal-2014, Hariharan-etal-2016, Carley-etal-2017}. However, significant uncertainties remain regarding the inferred magnetic field strengths and the detailed relationship between CMEs and type~IV radio bursts, particularly when relying solely on remote observational methods \citep{Morosan-etal-2019, Kumari-etal-2021}. 

Numerical models capable of realistically reproducing complex solar wind structures and simulating the acceleration and transport of energetic particles, combined with radio emission models, can help address open questions regarding the mechanisms underlying particle acceleration and radio emission. We demonstrate this approach here by coupling three numerical models to simulate GS emission from energetic electrons, including mildly relativistic and ultrarelativistic populations trapped within an erupting flux rope. We deploy the first-principles magnetohydrodynamical (MHD) coronal model COolfluid COroNal UnsTructured (COCONUT; \citealt{Perri-etal-2022}) constrained by observational magnetogram data at the inner boundary to generate realistic coronal conditions. This model represents the CME as an unstable Titov--Démoulin magnetic flux rope (MFR; \citealt{Titov-Demoulin-1999, Titov-etal-2014}). The transport of energetic electrons is modelled via the physics-based PArticle Radiation Asset Directed at Interplanetary Space Exploration (PARADISE; \citealt{Wijsen-etal-2019, Wijsen-2020}), which has been coupled to COCONUT previously \citep{Husidic-etal-2024}. The MHD parameters derived from COCONUT and electron distributions obtained from PARADISE are then passed into the Ultimate Fast Gyrosynchrotron Codes (UFGSCs; \citealt{Fleishman-Kuznetsov-2010, Kuznetsov-Fleishman-2021}) to derive synthetic type~IV radio spectra by computing the full GS emission and absorption coefficients using fast and accurate numerical approximations. By varying electron injection spectra and MFR configurations, and observing the evolving CME from different vantage points in the solar corona, we investigate how electron and CME properties influence the characteristics of radio spectra, including intensity, spectral shape, duration, and frequency range.

The remainder of the paper is structured as follows. In Sect.~\ref{sec:models}, we describe the coupled numerical models used to generate synthetic radio spectra. Simulation results are presented and discussed in Sect.~\ref{sec:simulations}, and we conclude with a summary and outlook in Sect.~\ref{sec:summary}. Additional technical details of the numerical models are provided in the Appendix~\ref{app: technical details}.

\section{Coupled numerical models and setup}\label{sec:models}
\begin{figure}
    \includegraphics[width=0.5\textwidth]{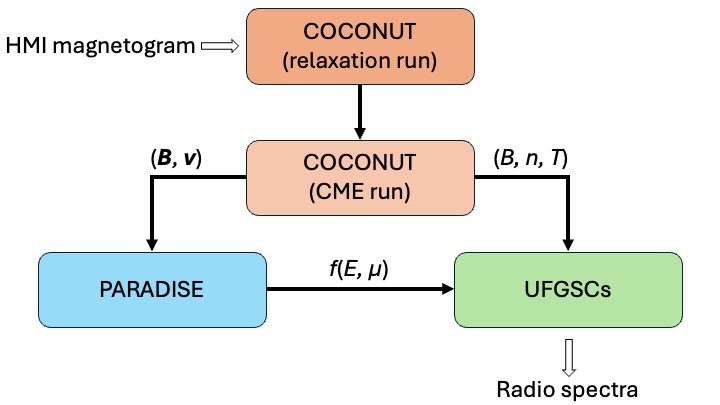}
   \caption{Illustration of the modelling chain for the generation of synthetic radio spectra.}
              \label{fig:model_chain}%
\end{figure}

Figure~\ref{fig:model_chain} illustrates the modelling chain to generate synthetic radio spectra, represented as a flowchart and described in the following subsections. The process begins by computing a relaxed corona using the MHD model COCONUT and an HMI magnetogram. The simulation is then restarted with a superimposed unstable MFR to represent the CME\footnote{hereafter, we refer to the MFR as the CME}. Next, the particle transport code PARADISE is used to evolve energetic electron distributions within the MHD environment. Finally, the synthetic GS emission is computed by combining the electron distributions from PARADISE and plasma parameters from COCONUT as input to the GS radiation code.

\subsection{The COCONUT model}\label{subsec:COCONUT}

COCONUT is a three-dimensional (3D) MHD model that solves the ideal MHD equations in a time-implicit manner, producing realistic coronal configurations. In this work, we use the recently updated version of COCONUT, which incorporates additional source terms for coronal heating, radiation losses, and thermal conduction into the energy conservation equation \citep{Baratashvili-etal-2024}. The generation of the coronal background proceeds in two stages. First, we compute the relaxed background solar wind using a magnetogram obtained from the Helioseismic and Magnetic Imager (HMI) onboard the Solar Dynamics Observatory, dated July 2, 2019. We then restart the COCONUT simulation, modifying only the magnetic field by superimposing an MFR at the solar surface. The magnetogram is taken from a period of solar minimum, providing a less complex coronal environment that minimises additional influences on the evolution of the CME and the energetic particle distribution.

The CME can be characterised by a dimensionless parameter $\zeta$. Generally, larger values of $\zeta$ correspond to more free magnetic energy, leading to more violent CMEs with higher injection speeds and stronger magnetic fields (see Appendix~\ref{app: COCONUT} for details). In this work, we use two setups: one with $\zeta = 30$ and one with $\zeta = 70$. The case $\zeta = 30$ produces a CME with an initial magnetic field strength of about 5.8~G and an initial speed of about $9.4\times 10^7$~cm/s, whereas the case $\zeta = 70$ results in a CME with a stronger field of about 10.6~G and a faster speed of roughly $1.3 \times 10^8$~cm/s. The remaining parameters for the two CMEs used in the Titov--D\'{e}moulin model are identical (see Appendix~\ref{app: COCONUT}).

\subsection{The PARADISE code}\label{subsec:PARADISE}

Next, the particle transport code PARADISE is employed to inject and propagate energetic electron distributions within the erupting MFR. The electron distributions evolve according to the focused transport equation \citep[FTE; e.g.,][]{vandenBerg-etal-2020}, which governs their anisotropic evolution by accounting for particle scattering, focusing, mirroring, and adiabatic deceleration, among other effects. Hence, PARADISE enables us to track the development of electron distributions in both pitch-angle and energy space along the evolving MFR, capturing the interplay between magnetic field variations and particle transport.

We select two different spectral indices for the electron injection spectra. Assuming a power-law energy distribution $\mathrm{d}N/\mathrm{d}E \propto E^{-\delta}$, we choose $\delta = 2$ and $\delta = 3$ with an initial energy range from 10~keV to 10~MeV and isotropic pitch-angle distribution. The electrons are injected as a delta function in time, approximately 30 minutes after the CME's eruption, within one of the CME flanks (i.e., the side regions of the MFR, curving downward from the apex) near its central axis and about $0.5~R_\odot$ above the solar surface. We adopt an injection method similar to that used by \cite{Husidic-etal-2024}, in which a specified high magnetic field value is used to identify the injection site. This approach has previously proven effective to ensure particle confinement within the CME.
The selected energy range is motivated by previous studies identifying 10~keV to 10~MeV as relevant for GS emission \citep{Dulk-Marsh-1982, Fleishman-Kuznetsov-2010, Kuznetsov-Fleishman-2021}, while spectral indices $\delta \ge 2$ are commonly observed (e.g., \citealp{Maksimovic-etal-1997}) and frequently adopted in GS calculations and modelling (e.g., \citealt{Dulk-1973, Dulk-Marsh-1982, Kuznetsov-Fleishman-2021}). At injection time, the CME reached a height of about 2.7~$R_\odot$ for $\zeta = 30$, and for $\zeta = 70$, approximately 3.9~$R_\odot$. By the end of the simulation, the corresponding CME heights were about 8.1~$R_\odot$ and 13.1~$R_\odot$, respectively.

The simulation results from PARADISE are expressed as differential intensities $j$, from which the particle distribution function $f$ can be derived according to
\begin{align}
    j(\mathbf{x},p, \mu,t) = p^2\,f(\mathbf{x},p, \mu,t)
\end{align}
with spatial vector $\mathbf{x}$, momentum magnitude $p$, pitch-angle cosine $\mu = \cos \alpha$ and pitch-angle $\alpha$, and time $t$. Because the FTE is linear, its solutions can be scaled by an arbitrary constant. In Appendix.~\ref{app: PARADISE} we describe the required units for the GS code and how they are obtained.

\subsection{The Ultimate Fast GS Codes}\label{subsec:GS code}

The UFGSCs contain a set of numerical schemes for calculating GS emission in the non-quantum regime, covering non-relativistic ($\gamma \approx 1$), mildly relativistic ($\gamma \gtrsim 1$), and ultrarelativistic ($\gamma \gg 1$) energy ranges \citep{Kuznetsov-Fleishman-2021}. While an earlier version of these GS codes was limited to using idealised (i.e., analytically defined) distribution functions in energy and pitch-angle cosine \citep{Fleishman-Kuznetsov-2010}, the latest update \citep{Kuznetsov-Fleishman-2021} allows the use of arbitrarily shaped electron distributions defined on a mesh. 

The UFGSCs were developed to circumvent the computationally demanding calculations of the exact equations for GS emission and absorption coefficients \cite{Melrose-1968, Ramaty-1969}. To achieve this, \cite{Fleishman-Kuznetsov-2010} refined the fast numerical approximations introduced by \cite{Petrosian-1981} and \cite{Klein-1987}, primarily by replacing the summation over cyclotron harmonics by integrating over a corresponding continuous parameter and including an approximate treatment of Bessel functions (for details, see \citealt{Fleishman-Kuznetsov-2010}). In the current version, the code can operate in one of three modes: purely exact, purely continuous, or hybrid. In the hybrid mode, the user specifies a threshold frequency $\nu_0 \equiv \eta \, \nu_\mathrm{g}$, where $\eta$ is a positive integer and $\nu_\mathrm{g} = e\,B / (2\,\pi\,m_\mathrm{e}\, c)$ denotes the electron gyrofrequency with elementary charge $e$, magnetic field magnitude $B$, and electron mass $m_\mathrm{e}$. Below this threshold, the code employs the exact mode, while above, it switches to the continuous mode. In this study, we use the UFGSCs in hybrid mode and set $\nu_0 = 12\, \nu_\mathrm{g}$ as a threshold for switching both to the continuous code and to the use of approximated Bessel functions, resulting in the code operating primarily in continuous mode, which smooths out harmonic structures and ensures a clean approximation of incoherent GS emission. While the UFGSCs also allow for the inclusion of bremsstrahlung from electron-ion (free-free) and electron-neutral (neutral bremsstrahlung) collisions, we disable those processes to isolate GS emission.

\subsection{Viewing geometry and data preparation}\label{subsec:GS code_setup}

To compute the GS emission from the COCONUT+PARADISE simulation, we define structured sets of lines of sight (LOS) and prepare the corresponding plasma input parameters at each discretised node, including the magnetic field magnitude $B$, the background solar wind number density $n$, the solar wind temperature $T$, and the viewing angle $\psi$ (i.e., the angle between a line of sight and the local magnetic field vector), along with the PARADISE electron distribution values. The UFGSCs independently calculate the emission and absorption coefficients along each line of sight. In many GS calculations, the emission is computed along a single line of sight and scaled by the assumed source area, which is sufficient for unresolved or compact sources. In our setup, the virtual spacecraft is located outside the MFR but still in the corona, so the emission must be computed over an extended field of view that captures the full spatial structure of the source. 

\begin{figure*}
    \centering
    \includegraphics[width=0.33\textwidth]{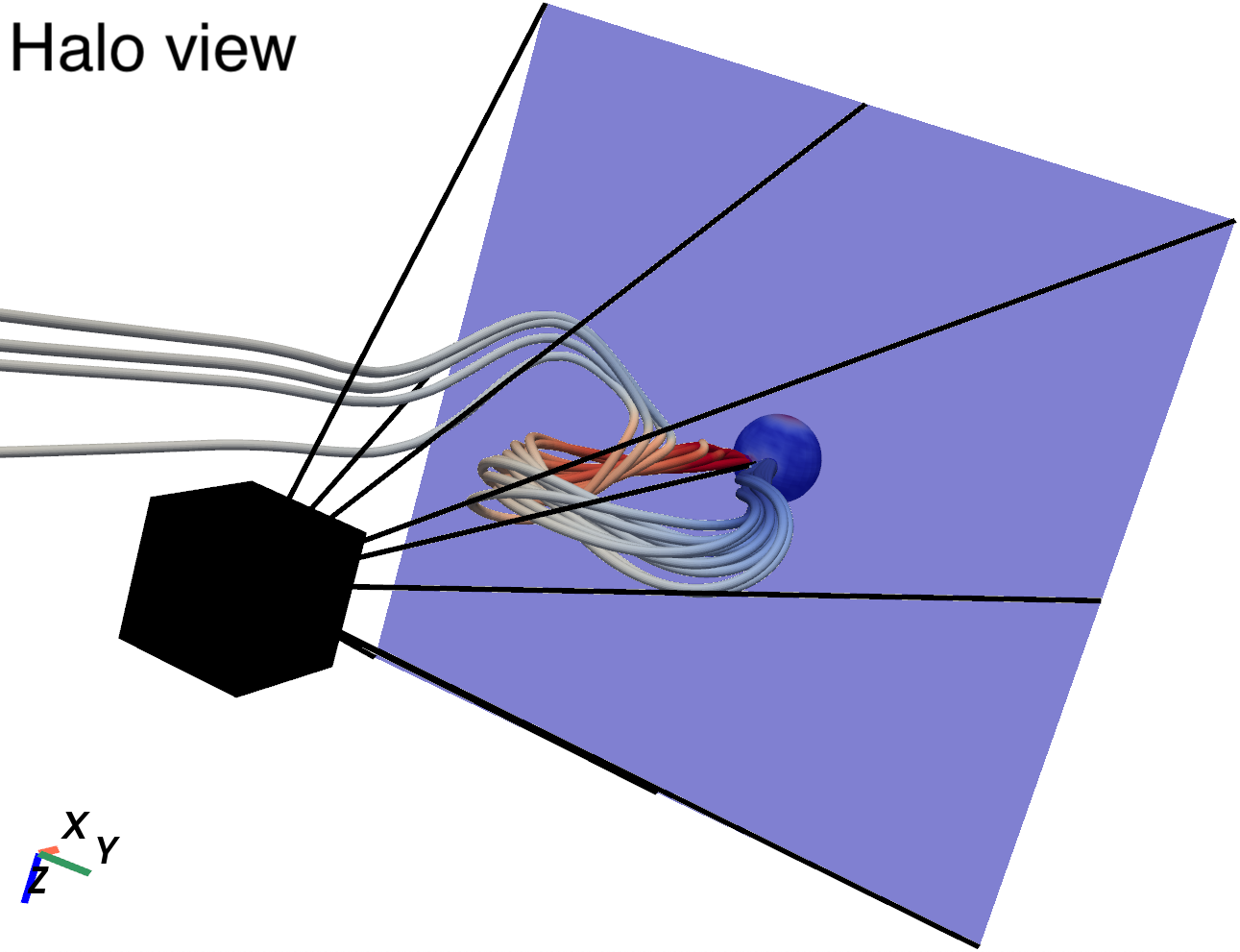}
    \includegraphics[width=0.33\textwidth]{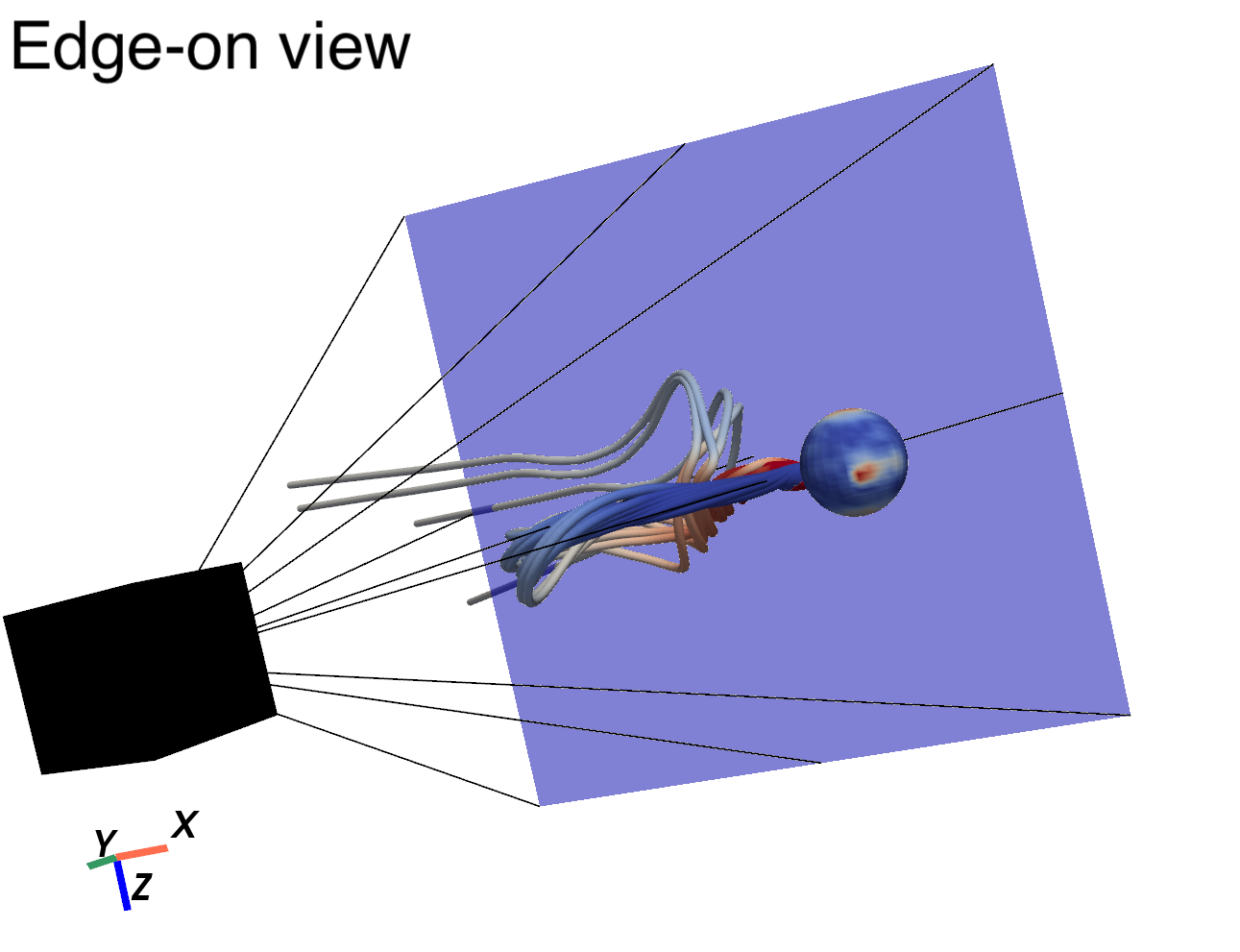}
    \includegraphics[width=0.33\textwidth]{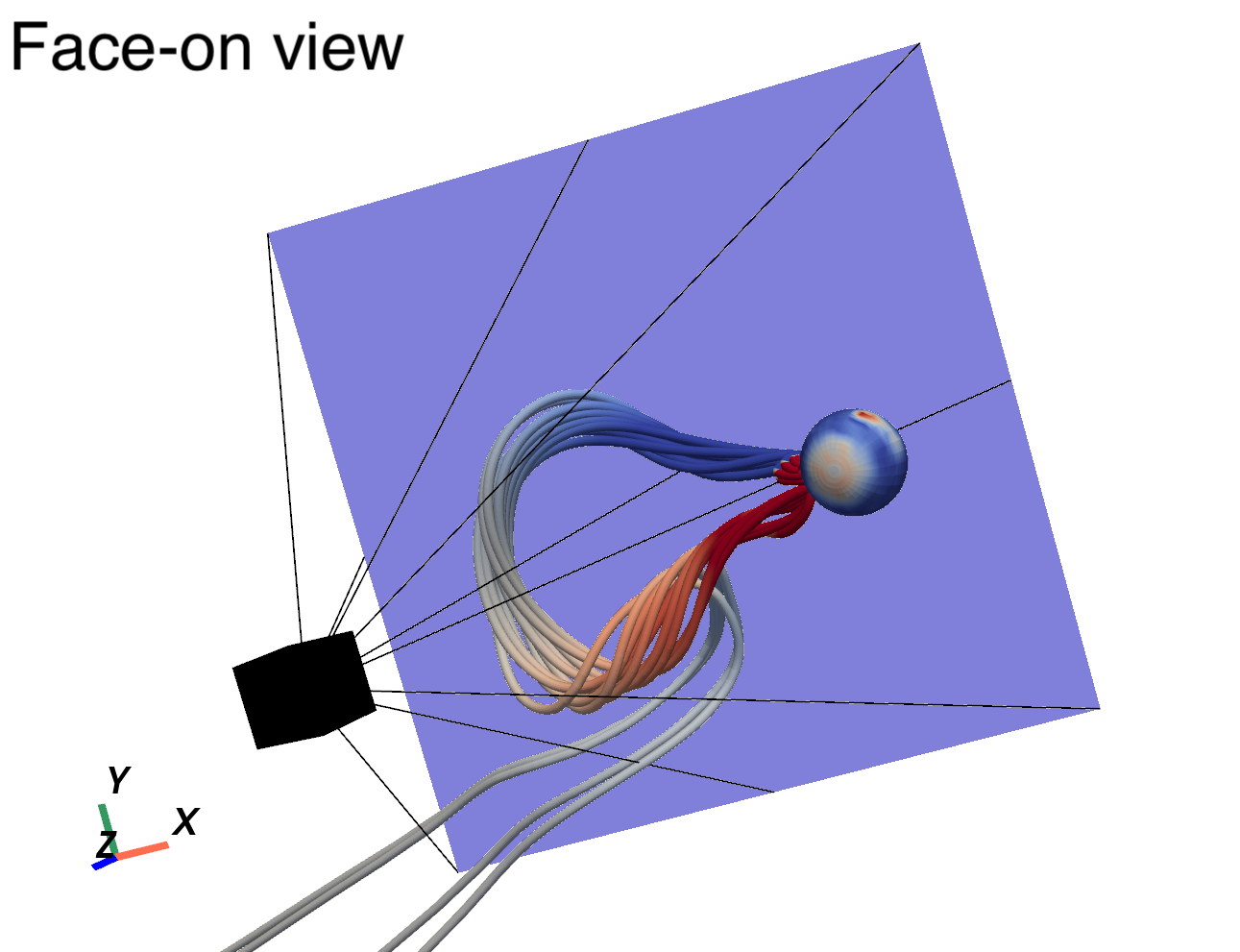}
    \caption{Positions of the virtual spacecraft and their respective viewing fields. The black cube represents the spacecraft, while the blue area illustrates the field of view. Some representative lines of sight are included.}
    \label{fig:viewing_fields}
\end{figure*}

For this study, we explore three distinct observational geometries by placing a virtual spacecraft at different locations relative to the erupting CME. Each spacecraft position is associated with a central line of sight perpendicular to a viewing field of $16\,R_\odot \times 16\,R_\odot$. The three viewing geometries are defined as follows: 
(1) the "halo view", where the CME propagates directly towards the observer -- a configuration typical of Earth-directed events and commonly observed as halo CMEs in coronagraph data (e.g., \citealt{Howard-etal-1982, Schwenn-etal-2005}); 
(2) the "edge-on view", where the observer is positioned to the side of the CME, with the central line of sight across the central axes of both flanks -- a geometry less commonly observed, as it would require the CME to be observed at the solar limb and tilted by approximately $90^\circ$; and
(3) the "face-on view", where the observer is positioned above the CME, viewing the full curvature of the flux rope --  a configuration commonly encountered when CMEs are observed near the solar limb (e.g., \citealt{Bastian-etal-2001, Kansabanik-etal-2024}). 
The coordinates of the spacecraft and base points of the central line of sight, along with the vantage points, CME and electron injection distribution parameters, are summarised in Table~\ref{tab:LOS}. Figure~\ref{fig:viewing_fields} illustrates these configurations, showing the virtual spacecraft (black cube), the CME, and the viewing field (blue-shaded area), along with some representative LOS. In all three cases, we use a resolution of $20 \times 20$ LOS, with each line of sight assigned an area of $A_\mathrm{total}/N_\mathrm{LOS}$, that is, the total viewing field area $A_\mathrm{total}$ is divided by the total number of LOS $N_\mathrm{LOS} = 400$.

\begin{table}
    \caption[]{Observer geometry and parameter combinations in the simulation setup.}
    \label{tab:LOS}
    $$
    \begin{array}{c|cc|c|c}
        \textnormal{Perspective} & \multicolumn{2}{c|}{\textnormal{Coordinates [$R_\odot$,$R_\odot$,$R_\odot$]}} & \zeta & \delta \\
        \cline{2-3}
        & \textnormal{Base point} & \textnormal{Spacecraft} & & \\
        \hline
        \text{halo view} & (-1.1,0,0) & (-18,0,0) & 30,\ 70 & 2,\ 3 \\
        \text{edge-on view} & (-8,-10,0) & (-3,14,0) & 30,\ 70 & 2,\ 3 \\
        \text{face-on view} & (-10,0,-12) & (-1.5,0,14) & 30,\ 70 & 2,\ 3
    \end{array}
    $$
\end{table}
\section{Simulation results and discussion}\label{sec:simulations}

\begin{figure*}
    \centering
    \begin{tabular}{cc} 
        \includegraphics[width=0.45\textwidth]{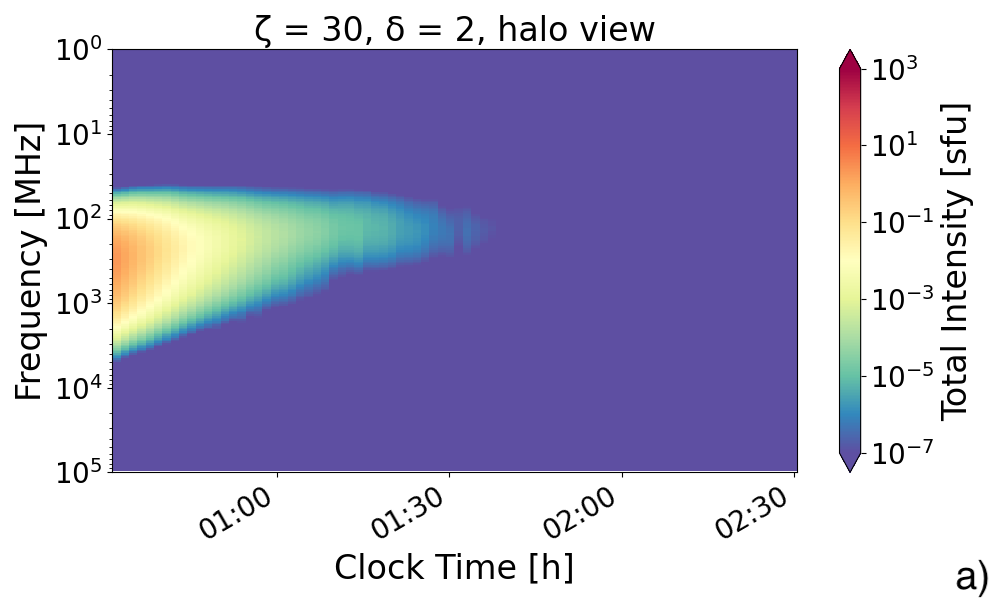} & 
        \includegraphics[width=0.45\textwidth]{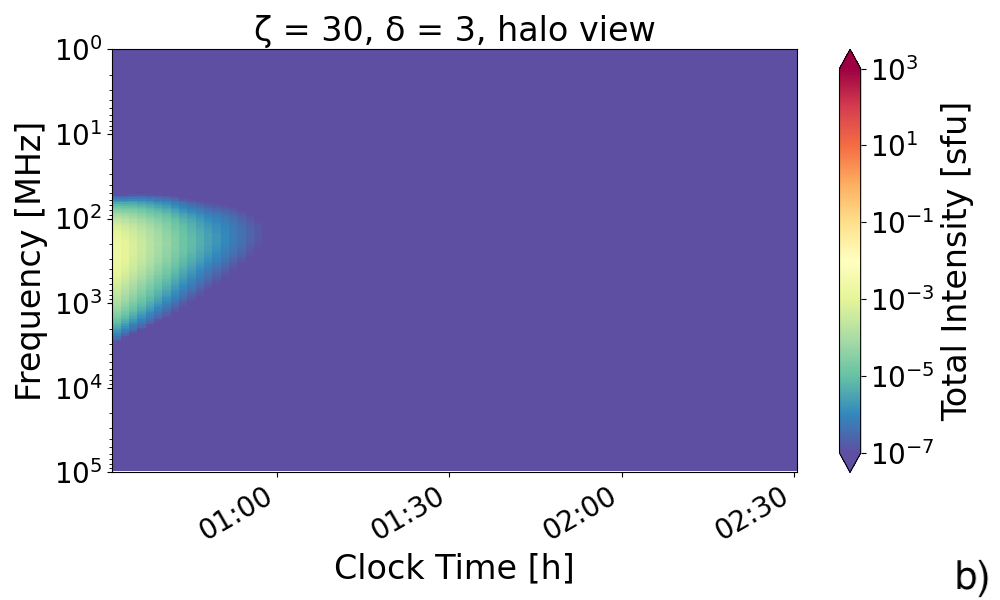} \\
        \includegraphics[width=0.45\textwidth]{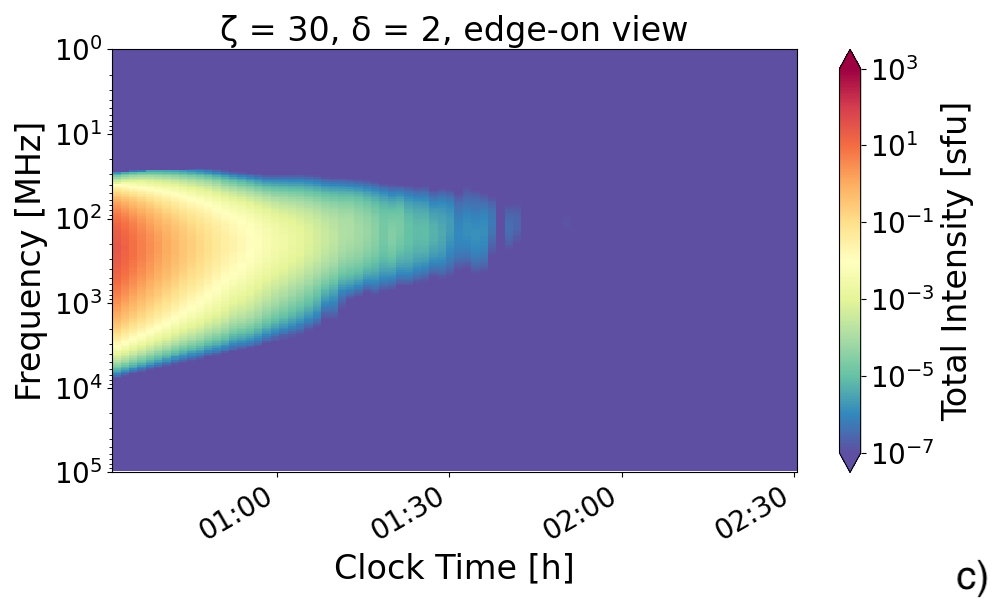} & 
        \includegraphics[width=0.45\textwidth]{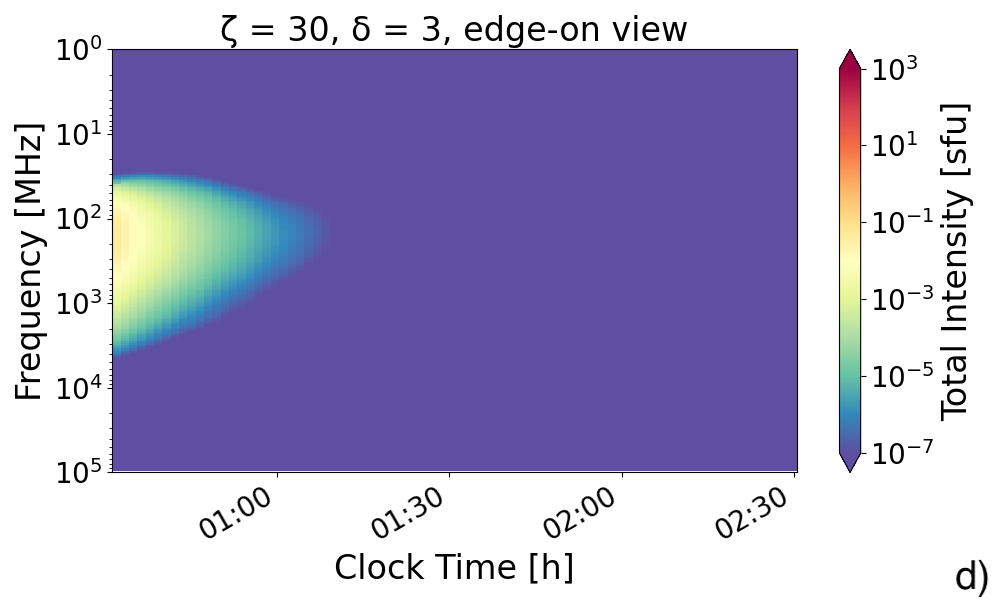} \\
        \includegraphics[width=0.45\textwidth]{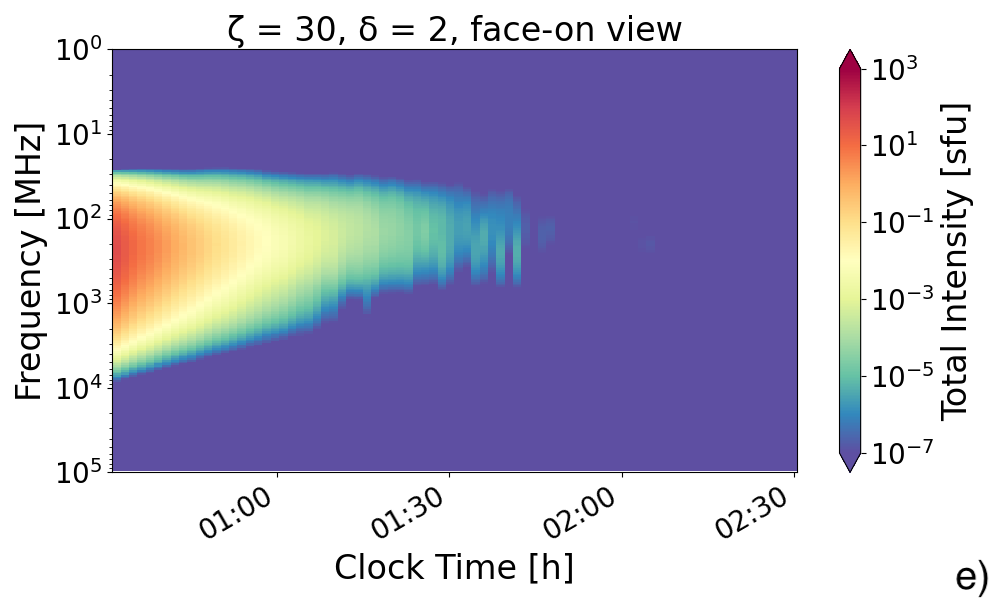} & 
        \includegraphics[width=0.45\textwidth]{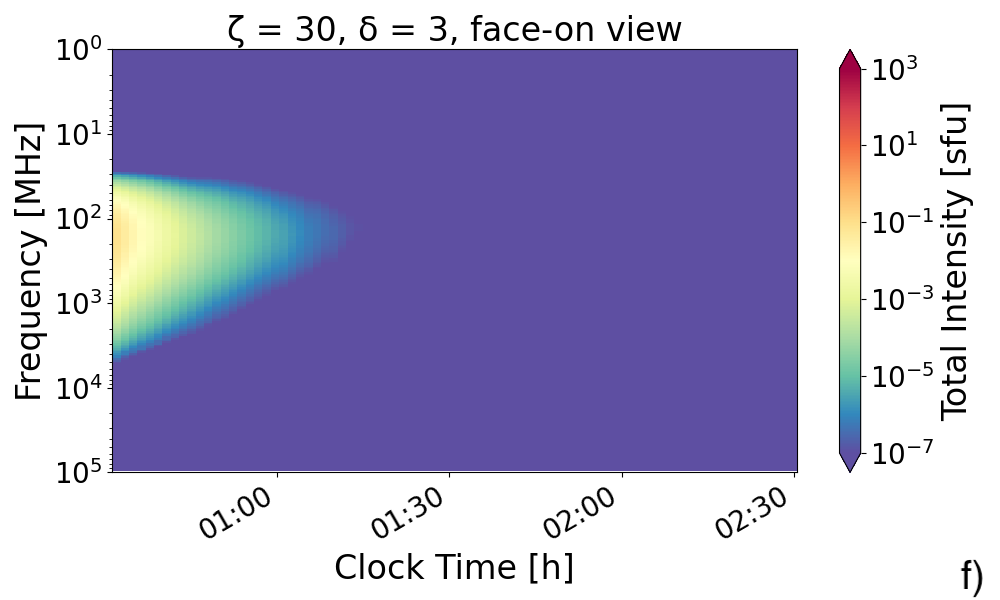} \\
    \end{tabular}
    \caption{Synthetic radio spectra for the CME case with $\zeta = 30$. The panels in the left column show the observed radio emission from the injection spectrum with $\delta = 2$ seen from the three spacecraft locations, while the panels in the right column contain the radio emission from the injection spectrum with $\delta = 3$.}
    \label{fig:radio_spectra_zeta_30}
\end{figure*}
\begin{figure*}
    \centering
    \begin{tabular}{cc} 
        \includegraphics[width=0.45\textwidth]{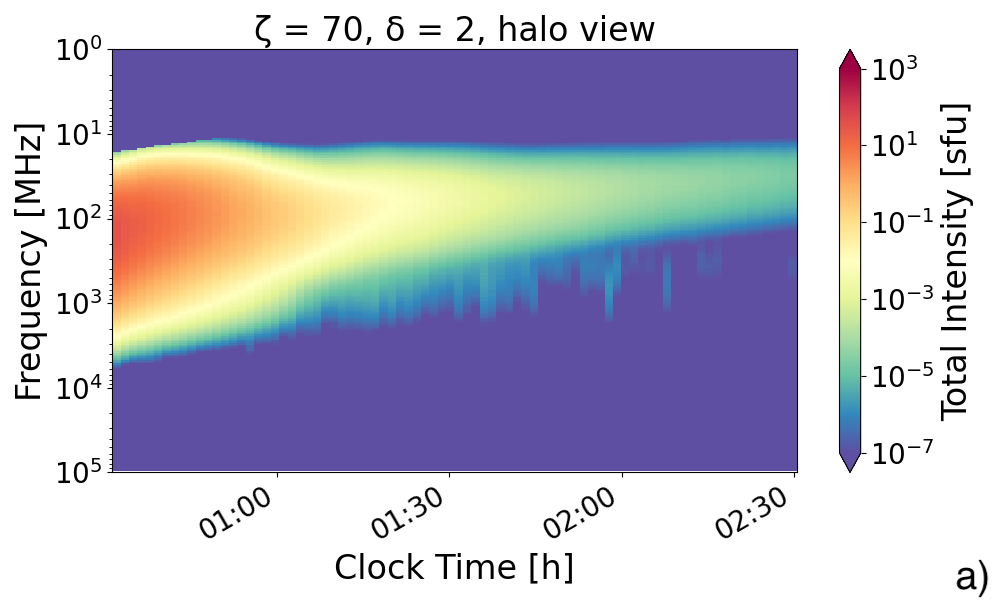} & 
        \includegraphics[width=0.45\textwidth]{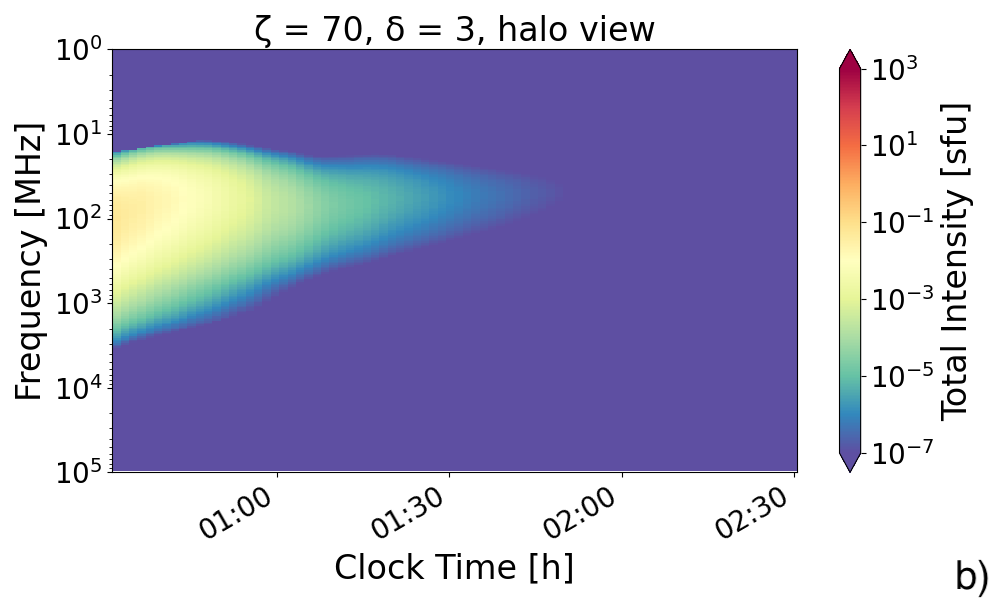} \\
        \includegraphics[width=0.45\textwidth]{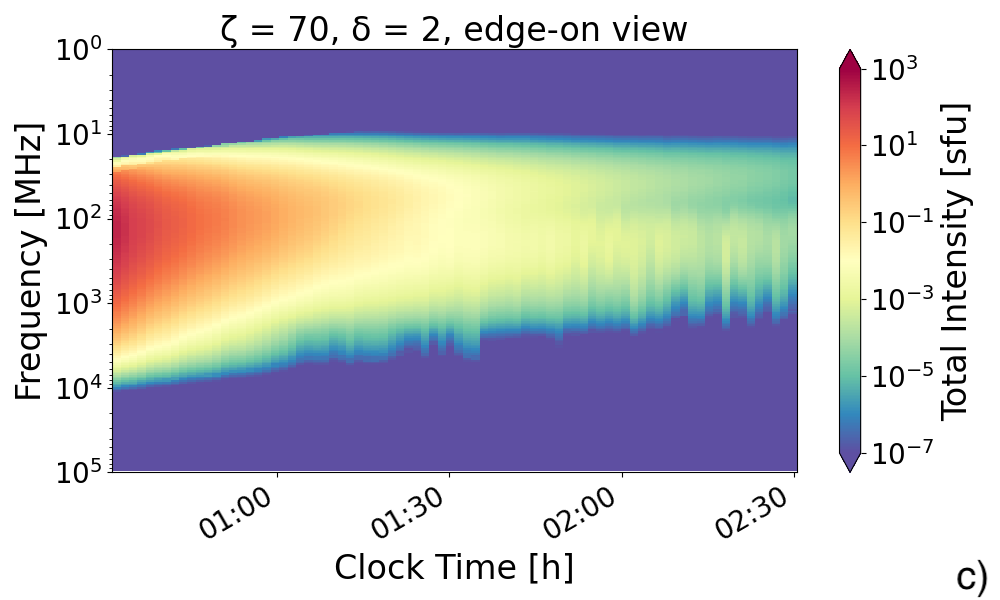} & 
        \includegraphics[width=0.45\textwidth]{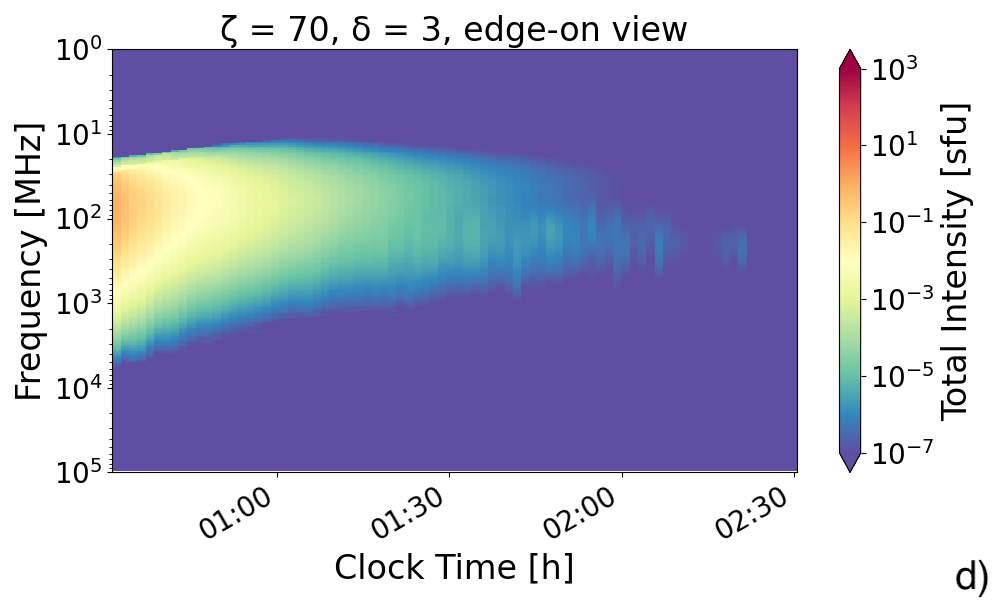} \\
        \includegraphics[width=0.45\textwidth]{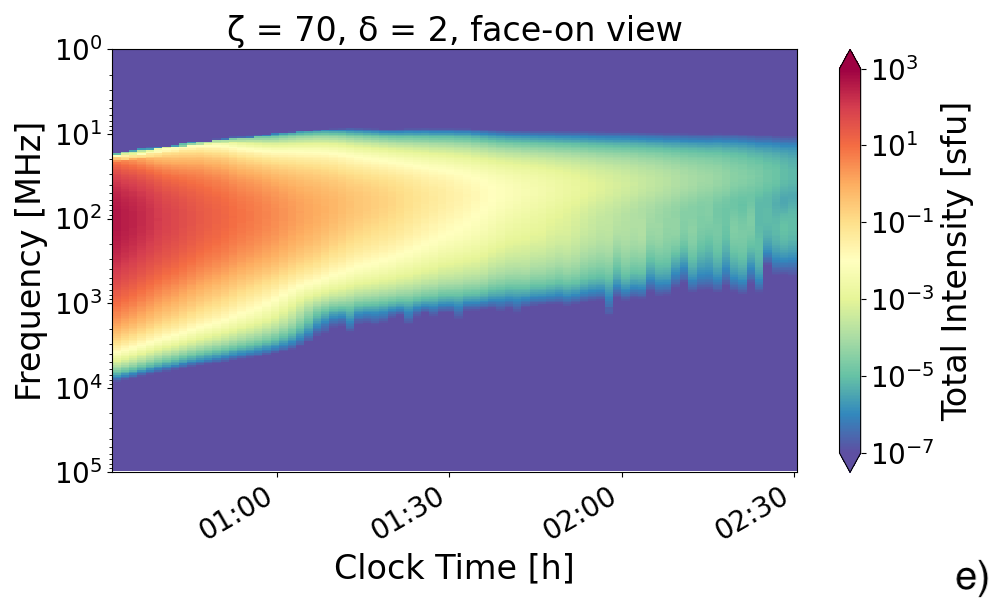} & 
        \includegraphics[width=0.45\textwidth]{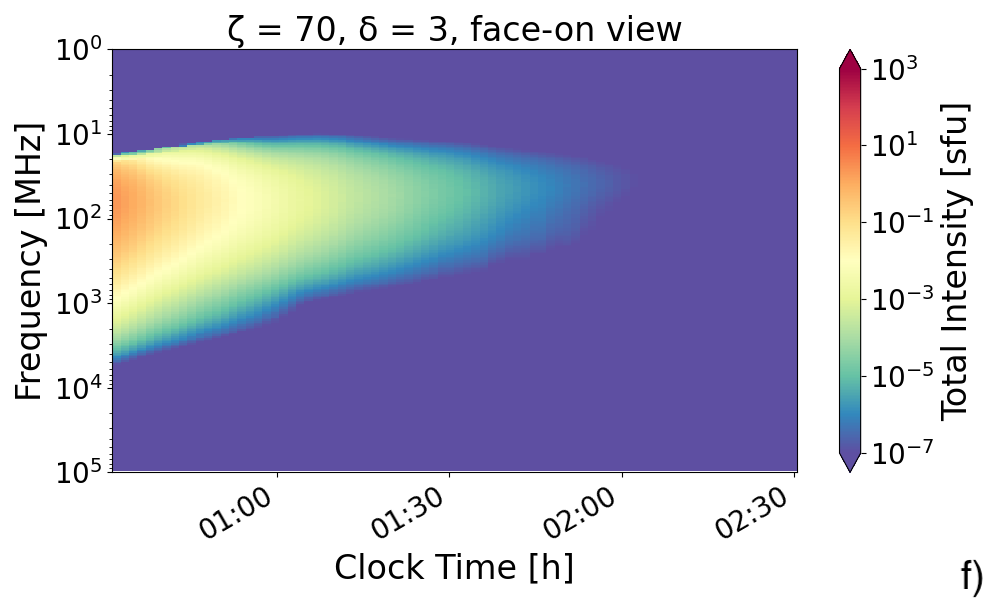} \\
    \end{tabular}
    \caption{Synthetic radio spectra for the CME case with $\zeta = 70$. The panels in the left column show the observed radio emission from the injection spectrum with $\delta = 2$ seen from the three spacecraft locations, while the panels in the right column contain the radio emission from the injection spectrum with $\delta = 3$.}
    \label{fig:radio_spectra_zeta_70}
\end{figure*}

We investigate the impact of different CME configurations and energy injection spectra on the resulting GS emission. Two CME setups are considered, classified by the $\zeta$-parameter with values 30 and 70, and each is paired with injected electron energy distributions of spectral index $\delta = 2$ and $\delta = 3$. This section refers to individual cases by their respective $\zeta$- and initial $\delta$-values. However, the actual spectral indices of the electron populations vary over time and location due to transport effects, such as adiabatic cooling, particle losses, and scattering.

We begin by outlining the general features shared across all computed radio spectra and how they compare to real observations of type~IV bursts. All obtained radio spectra (Figs.~\ref{fig:radio_spectra_zeta_30} and \ref{fig:radio_spectra_zeta_70}) exhibit a similar structure, containing a high-intensity centre surrounded by weaker emission. This feature is more pronounced in the $\delta = 2$ cases compared to those with $\delta = 3$. Similar intensity structures are regularly observed in dynamic spectra of type~IV bursts, as illustrated in Fig.~\ref{fig:observational_data} and documented by studies such as \cite{Melnik-etal-2018, Vasanth-etal-2019, Morosan-etal-2020, Kumari-2022}. The high-intensity cores in our simulations emerge from the solution due to the local electron distribution, the CME magnetic field strength, and the viewing geometry (see paragraphs below). Broadband features with locally spiky or burst-like enhancements are often associated with emission originating in the MFR flanks near the footpoints, where magnetic field strengths are high and pitch-angle anisotropies are pronounced. These conditions are entirely consistent with the GS framework.

Furthermore, all twelve spectra exhibit a drift of the intensity peak from higher to lower frequencies over time, resulting in broad, upward-drifting emission lanes in the time-frequency domain (with the frequency axis being inverted in the plots). This behaviour is typical for synchrotron emission from electrons in the energy range of 1 -- 10~MeV in expanding, twisted MFRs, such as the Titov--D\'{e}moulin model used in our simulations, where the internal magnetic field weakens with time. Since the synchrotron peak frequency scales approximately with the square of the magnetic field strength ($\nu_\mathrm{peak} \propto B^2$), this leads to a downward shift in frequency as the CME evolves. This effect is reinforced by the adiabatic cooling of electrons, which reduces their energy as they propagate. To quantify this behaviour, Table~\ref{tab:drift_rates} lists estimated drift rates (in MHz/s) for all twelve spectra shown in Figs.~\ref{fig:radio_spectra_zeta_30} and \ref{fig:radio_spectra_zeta_70}. Drift rates were obtained by applying a linear fit to the time evolution of the intensity peaks over the first $\sim$~30~min (in order to compare across all cases). While no clear trend emerges across all cases, we find that the halo-view configuration consistently exhibits the strongest negative drift rates.

Frequency drifts in type~IV spectra, especially those associated with CMEs, have been frequently reported across a broad range of values. For instance, \cite{Kumari-etal-2021} identified downward frequency drifts in the majority of CMEs in their sample, using drift rates (in absolute value) $\vert R_\mathrm{drift}\vert \ge 0.03$~MHz/s as an indicator for IVm bursts. The authors found that most type~IV bursts exhibited drift rates $\vert R_\mathrm{drift}\vert \le 0.5$~MHz/s. Other studies report higher drift rates for fine structures embedded within type~IV continua, ranging from several MHz/s up to tens of MHz/s \citep{Nishimura-etal-2013, Melnik-etal-2018}, while the background continuum typically drifts more slowly, with $\vert R_\mathrm{drift}\vert \approx 10$~kHz/s \citep{Melnik-etal-2018}. In this context, the drift rates obtained from our simulations fall towards the lower end of the reported literature values. A plausible explanation may lie in our modelling framework including only synchrotron/GS emission, where frequency drifts primarily arise from the gradual weakening of the magnetic field within the expanding MFR. It should be further noted that the reported drift rates were mostly derived at lower frequencies compared to our results, which may additionally explain discrepancies.

The GS emission is detected at all spacecraft positions from the time of electron injection (approximately 30~min into the simulation, with $t = 0$~min being the time of CME eruption), as the CME is initially entirely within the field of view. In general, the radio bursts are obtained in frequency ranges between 20~MHz and 8~GHz, with core intensities $> 10^{-4}$~sfu (solar flux units) and $< 10^3$~sfu (see peak intensities in Table~\ref{tab:drift_rates}), and durations between $\sim 30$~min and $\sim 2$~h.

While type~IV bursts are commonly detected in the $10^1$--$10^2$~MHz range \citep{Melnik-etal-2018, Vasanth-etal-2019, Morosan-etal-2020, Kumari-etal-2021, Kumari-2022, Mohan-etal-2024}, observations also show extensions to higher frequencies from $10^2$--$10^3$~MHz \citep{Liu-etal-2018, Morosan-etal-2019} to several $10^3$~MHz \citep{Xie-etal-2002, Nishimura-etal-2013, Karlicky-Rybal-2020}. Similarly, reported spectral flux densities span a wide range, from $10^1$--$10^2$~sfu \citep{Dulk-1973, Carley-etal-2017, Melnik-etal-2018} over multiple $10^2$~sfu \citep{Bouratzis-etal-2015} to around $10^3$~sfu \citep{Melnik-etal-2018}, and even approaching $10^4$~sfu in cases involving fine structures \citep{Alissandrakis-etal-2019}. 
It is important to note that the peak intensities also depend on the number of injected electrons.
Finally, recorded type~IV durations are also highly variable, ranging from a few minutes \citep{Karlicky-Rybal-2020, Morosan-etal-2021, Kumari-etal-2021, Mohan-etal-2024} to several hours \citep{Liu-etal-2018, Melnik-etal-2018, Vasanth-etal-2019}. Historically, type~IVm were associated with durations $< 1$~h, while durations $>1$~h were attributed to type~IVs. However, more recent studies report IVm events lasting $>1.5$~h \citep{Ramesh-etal-2013} and even $>2.5$~h (\citealt{Vasanth-etal-2019}; see also the discussion in \citealt{Kumari-etal-2021}). These comparisons suggest that our synthetic GS spectra are physically realistic in terms of frequency range, spectral flux density, and duration.

When comparing the intensities for different spectral indices of the injected electrons at each spacecraft location, we consistently find higher intensities and longer radio burst durations for $\delta = 2$ compared to $\delta = 3$. This outcome is expected, as the spectrum with $\delta = 2$ is flatter than that with $\delta = 3$, meaning a larger fraction of the electron population is in the high-energy range. Consequently, the GS spectrum is dominated by higher-energy electrons, which produce stronger emission due to their greater radiative power (see Li\'{e}nard formula or Eq.~\ref{eq:P_syn}). We further note that, due to the energy dependence of particle transport, the intensity ratios between these two cases are not constant over time. The case with initial $\delta = 2$ not only contains more electrons of higher energy, but also a greater fraction of faster particles, yielding different transport behaviour over time and leading to distinct evolution of the injection spectra.

Next, we analyse the individual CME cases. Figure~\ref{fig:radio_spectra_zeta_30} presents the radio spectra obtained for the case $\zeta = 30$, with $\delta = 2$ in the left column and $\delta = 3$ in the right column. The spectra are shown from the point of particle injection up to about 2.5~hours into the simulation. Considering the left column ($\delta = 2$), at the first spacecraft position, where the observer is directly ahead of the CME, we find a peak intensity ratio of approximately $I_{\mathrm{max},\delta=2}/I_{\mathrm{max},\delta=3} \approx 1100$. At the second spacecraft location (observer viewing the MFR edge-on), the ratio is about 564. A similar ratio (roughly 560) is observed, when the observer is directly above the CME (face-on view). The results are qualitatively similar for the $\delta = 3$ cases in the right column (see Table~\ref{tab:intensity_ratios}).

The synthetic GS emission across different spacecraft positions also reveals noteworthy differences. Comparison of peak intensities between the edge-on and halo view for $\zeta = 30$ and $\delta = 2$ (panels a and c in Fig.~\ref{fig:radio_spectra_zeta_30}) indicates a ratio of $I_{\mathrm{max},\mathrm{edge-on}}/I_{\mathrm{max},\mathrm{halo}} \approx 13$. In contrast, a comparison between the face-on and halo view (panels a and e) yields $I_{\mathrm{max},\mathrm{face-on}}/I_{\mathrm{max},\mathrm{halo}} \approx 23$, highlighting the strong variation in GS intensity with observer vantage point.

The variation in the radio emission spectra across the three observer positions can be understood in terms of the local magnetic field strength and the viewing angle. The GS emission is sensitive to the observer's viewpoint relative to the MFR geometry. The GS brightness is proportional to the component of the magnetic field perpendicular to the electron motion, given by the electron pitch-angle, as $B_\perp = B\,\sin\alpha$. Thus, the viewing angle has a strong influence on both the emission intensity and the visibility of frequency drift. 

The halo view aligns the observer's LOS with the magnetic field lines that are essentially approximately parallel to the viewing direction, particularly along the MFR flanks. The field lines become more perpendicular only closer to the CME apex. Because GS emission is strongest when the magnetic field is perpendicular to the LOS, this geometry results in weaker overall emission, although the frequency drift remains visible.

From the edge-on view, the observer's LOS traverse both CME flanks, encountering a thicker cross-section of the MFR where magnetic fields are both stronger and more perpendicular to the LOS. This configuration enhances the GS brightness and supports magnetic mirroring, as the magnetic field converges near the anchored MFR footpoints, which remain part of the closed field structure and exhibit the highest magnetic field strengths in the simulation. These conditions enable effective particle trapping and sustained GS emission.

Finally, the face-on view provides visibility of both CME flanks and part of the apex, with many LOS quasi-perpendicular to the flux rope axis. This geometry yields the strongest GS emission of the three configurations, owing to both favourable pitch-angle orientation and a large effective emitting region. Again, qualitatively similar results were obtained for the simulations with spectral index $\delta = 3$ (see Table~\ref{tab:intensity_ratios}). It is worth noting that these results are specific to the generated MFR configuration. Increasing the twist of the MFR would modify the magnetic field geometry, change local viewing angles, and thus modify the observed GS emission.

Figure~\ref{fig:radio_spectra_zeta_70} shows the radio spectra obtained from the COCONUT simulation with the CME of $\zeta = 70$. The plots are arranged similarly to those in Fig.~\ref{fig:radio_spectra_zeta_30}, that is, the left column presents results based on spectral index $\delta = 2$, while the right column shows the plots for $\delta = 3$. Comparing the results for different spectral indices and spacecraft positions, we find qualitatively the same trends as in the CME case $\zeta = 30$. Again considering the $\delta = 2$ spectrum, quantitatively, detecting GS emission from the halo view, the ratio between the peak intensities for the two spectral indices is approximately $I_{\mathrm{max},\delta=2}/I_{\mathrm{max},\delta=3} \approx 620$. At the edge-on-view location, we find a ratio of approximately 310, while at the face-on-view position, a ratio of about 208 is observed. The $\delta = 3$ electron spectrum results are qualitatively similar (see Table~\ref{tab:intensity_ratios}). The differences across observer positions again indicate that from the face-on-view vantage point, the highest intensity is detected, followed by the edge-on-view and halo-view positions. The ratios between the peak intensities are $I_{\mathrm{max},\mathrm{edge-on}}/I_{\mathrm{max},\mathrm{halo}} \approx 7.2$ and $I_{\mathrm{max},\mathrm{face-on}}/I_{\mathrm{max},\mathrm{halo}} \approx 12.8$. 

In the $\zeta = 70$ case, our simulations additionally reveal a secondary emission lane at higher frequencies, distinct in drift rate and typically weaker than the main lane. These secondary lanes appear only in the edge-on and face-on view configurations, where some of the LOS from the observer's viewing field intersect regions in the lower MFR flanks close to the MFR footpoints. Although the actual footpoints are not directly visible in these views, the detected GS emission originates from nearby locations where the magnetic field is strong, and the LOS orientation is largely perpendicular to the local field direction. These regions also exhibit enhanced pitch-angle anisotropies and may contain distinct populations of high-energy electrons. 
The footpoints themselves remain anchored and largely static; however, the close MFR flank regions continue to expand over time, leading to moderate field weakening and a slight frequency drift.
As electrons move into stronger magnetic fields near the footpoint regions, they conserve their magnetic moment $\mu_\mathrm{m} = p^2_\perp/(2\,m_\mathrm{e}\,B)$ with perpendicular momentum component $p_\perp$, requiring an increase in pitch-angle and leading to magnetic mirroring. This process results in localised, time-varying enhancements in GS brightness that manifest as the observed secondary emission lanes. 

The total synchrotron power from a single relativistic electron can be derived from the Li\'{e}nard formula \citep{Rybicki-Lightman-1986}, and is approximately given by 
\begin{align}
P_{\mathrm{syn}} \approx \frac{4}{3}\,\sigma_\mathrm{T}\,c\,\gamma^2\,\beta^2\,U_B\,\sin^2\alpha\,, \label{eq:P_syn}
\end{align}
where the emission strength depends on the electron speed, the magnetic energy density $U_B = B^2/(8\,\pi)$, and the pitch-angle term $\sin^2\alpha$. Here, $\sigma_\mathrm{T}$ denotes the Thomson scattering cross-section, $c$ the speed of light in vacuum, and $\beta = v/c$. Near the MFR footpoints, both $B$ and $\alpha$ tend to be large, substantially enhancing $P_{\mathrm{syn}}$. As electrons are trapped within the MFR, they mirror at the stronger magnetic fields near the footpoints and travel back toward the weaker field near the MFR apex. Each time they pass through the footpoint regions, they produce bursts of local synchrotron emission, leading to a periodic emission pattern.
Since the lower MFR regions near the magnetic footpoints are compact and maintain strong magnetic fields and large pitch angles, they produce locally strong GS emission despite their limited spatial extent. In our simulation setup, electrons are magnetically confined and mirror in these lower regions of the MFR, which remain anchored to the surface. This trapping mechanism contributes to the intermittent GS enhancements observed along the LOS intersecting these regions.

In contrast, the primary GS lane originating from the expanding MFR's main body appears smoother and more continuous. The gradual expansion of the MFR causes the magnetic field magnitude to decrease, leading to a steady, upward frequency drift. Although magnetic fields near the footpoints are stronger, the total emission is dominated by the larger electron population distributed throughout the main body of the flux rope. The secondary emission lane, by comparison, appears more intermittent and consists of faint, locally enhanced features of GS emission. These features arise from a smaller electron population and are linked to confined mirroring in lower regions of the MFR flanks near the footpoints.

In more complex scenarios, flux ropes can expand rapidly and non-ideally, in the solar corona developing kinks, deflections, and asymmetries. These deformations may produce multiple distinct drifting GS lanes, each dependent on observer perspective and characterised by unique emission signatures. While our model assumes a simplified scenario with constant footpoints and no new flux emergence or footpoint diffusion, more intricate flux rope dynamics, such as increased twist or enhanced poloidal magnetic fields, could introduce localised gradients that might alter electron mirroring. In such cases, bursty broadband emissions may arise near footpoints and throughout other regions of the flux rope structure. 

We note that the employed MFR model is already significantly more realistic than many idealised models historically used in GS studies (e.g., \citealt{Dulk-1973, Fleishman-Melnikov-2003, Kuznetsov-etal-2015}). Our erupting MFR introduces a dynamically evolving magnetic topology with strong twist, leading to varying field curvatures and strengths, all of which directly affect particle transport and confinement. Furthermore, the continued expansion and radial outward propagation of the MFR also involves time-dependent variations in both field strength and geometry, capturing important physical aspects relevant to GS emission in a type~IV bursts. In contrast, flux emergence and footpoint diffusion processes occur on much longer timescales (e.g., \citealt{Giacalone-Jokipii-2004}) than those associated with the particle acceleration and GS emission modelled here in the low corona, and are therefore expected to have negligible influence on the simulation results.

Finally, we may compare the differences in the intensities of the GS emission for the different CME cases. Considering the spectral index $\delta = 2$, we note from the halo view observer that the peak intensity in the $\zeta = 70$ case is about 16 times higher compared to $\zeta = 30$, at the edge-on view about 9 times higher, and at the face-on view also about 9 times higher. Qualitatively similar results are found for the $\delta = 3$ spectra, and all ratios are summarised in Table~\ref{tab:intensity_ratios}. The stronger GS intensities in the $\zeta = 70$ simulation can be attributed to the stronger magnetic field strengths present within the CME, consistent with Eq.~\eqref{eq:P_syn}. Additionally, a higher degree in twist associated with larger $\zeta$-values may enhance the variation in the magnetic field along the MFR, promoting more effective magnetic mirroring and particle confinement, thus yielding longer-lasting and more intense GS emission.

\begin{table}[ht]
\centering
\caption{Frequency drift rates $R_\mathrm{drift}$ estimated over the first $\sim 30$~minutes after electron injection, and peak intensities for the different simulation configurations.}
\begingroup
\begin{tabular}{ccrcr}
\hline
$\delta$ & $\zeta$ & View & $R_\mathrm{drift}$ [MHz/s] & $I_\mathrm{max}$ [sfu] \\
\hline
2 & 30 & Halo & -0.0875 & $\sim$ 2\\
2 & 30 & Edge-on & -0.0199 & $\sim$ 28\\
2 & 30 & Face-on & -0.0251 & $\sim$ 48\\
3 & 30 & Halo & -0.0614 & $\sim$ 0.002\\
3 & 30 & Edge-on & -0.0064 & $\sim$ 0.05\\
3 & 30 & Face-on & -0.0025 & $\sim$ 0.09\\
2 & 70 & Halo & -0.0372 & $\sim$ 35\\
2 & 70 & Edge-on & -0.0271 & $\sim$ 248\\
2 & 70 & Face-on & -0.0075 & $\sim$ 442\\
3 & 70 & Halo & -0.0192 & $\sim$ 0.06\\
3 & 70 & Edge-on & -0.0051 & $\sim$ 0.8\\
3 & 70 & Face-on & -0.0010 & $\sim$ 2\\
\hline
\end{tabular}
\endgroup
\label{tab:drift_rates}
\end{table}
\begin{figure}
    \includegraphics[width=0.5\textwidth]{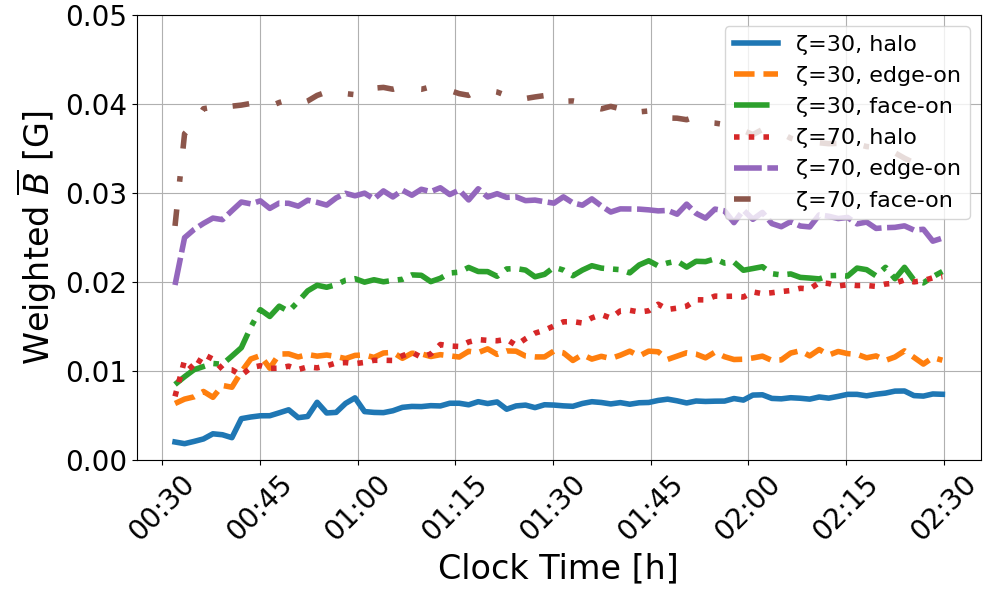}
   \caption{Comparing the weighted mean magnetic field strength over time for different observer positions and CME cases. For calculating the weighted average $\overline{B}$, we used the electron distribution of the $\delta = 2$ case. Each curve corresponds to a specific vantage point (halo, edge-on, face-on) and CME case ($\zeta = 30$, $\zeta = 70$).}
              \label{fig:B_field}%
\end{figure}

To further illustrate the differences in the observed GS emission intensities and their relationship to the magnetic field configuration, Fig.~\ref{fig:B_field} shows the weighted mean magnetic field strength evolution for different observer vantage points and CME setups. 
The mean field was computed by weighting the magnetic field along each line of sight using the electron distribution from the $\delta = 2$ case, followed by averaging across all LOS for each observer position. These values are not direct measurements of the local field at emission sites, but rather effective field strengths contributing to the observed GS signal from each vantage point. As a result, they tend to be lower than typical observational estimates (e.g., \citealt{Maia-etal-2007, Tun-Vourlidas-2013}), which often reflect localised peak values near the emission core. In our model, localised fields in parts of the CME (e.g., in the MFR flanks) are indeed higher, but their contribution is diluted in the averaged LOS-weighted magnetic field.

\begin{figure*}
    \centering
    \begin{tabular}{cc} 
        \includegraphics[width=0.45\textwidth]{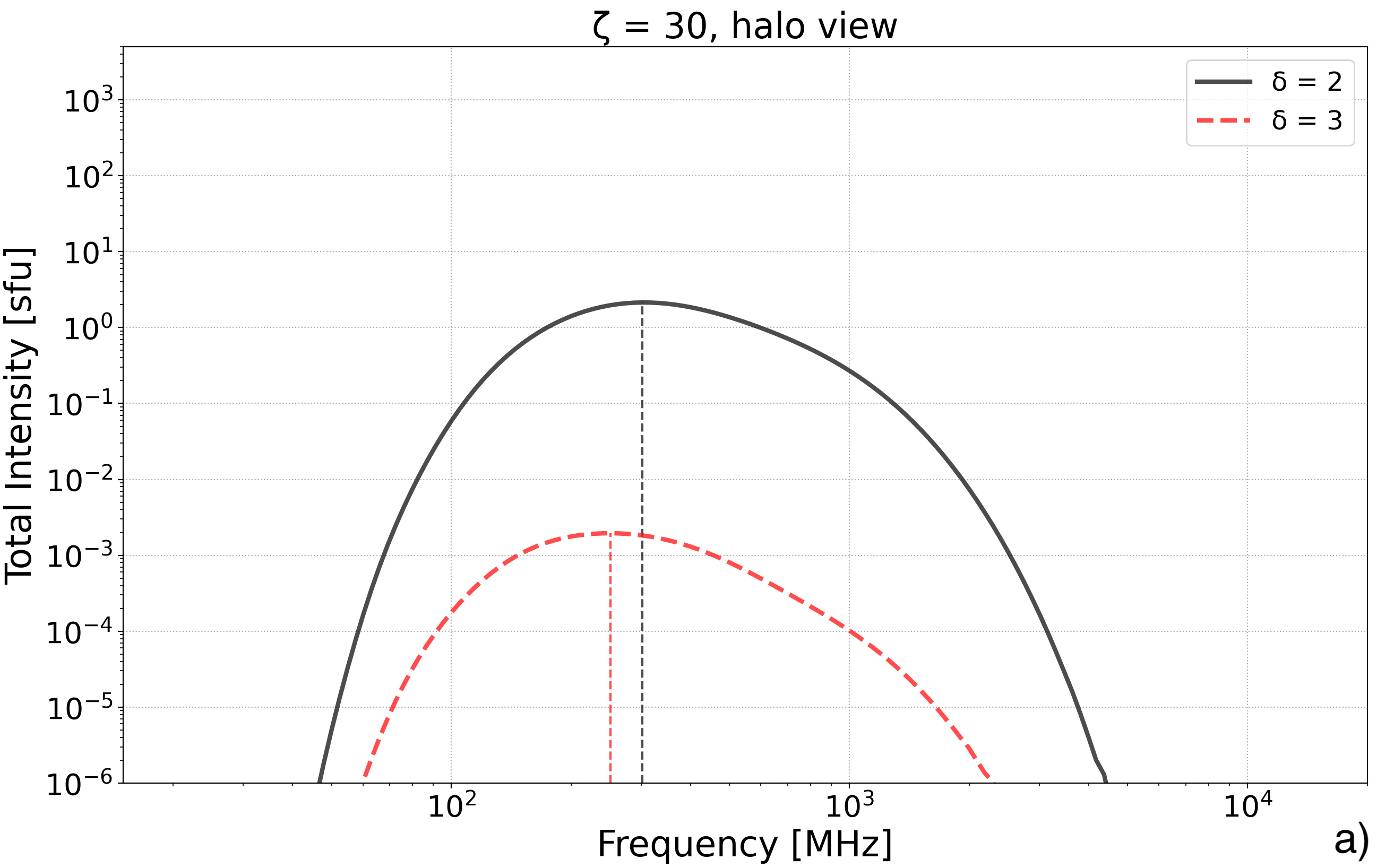} & 
        \includegraphics[width=0.45\textwidth]{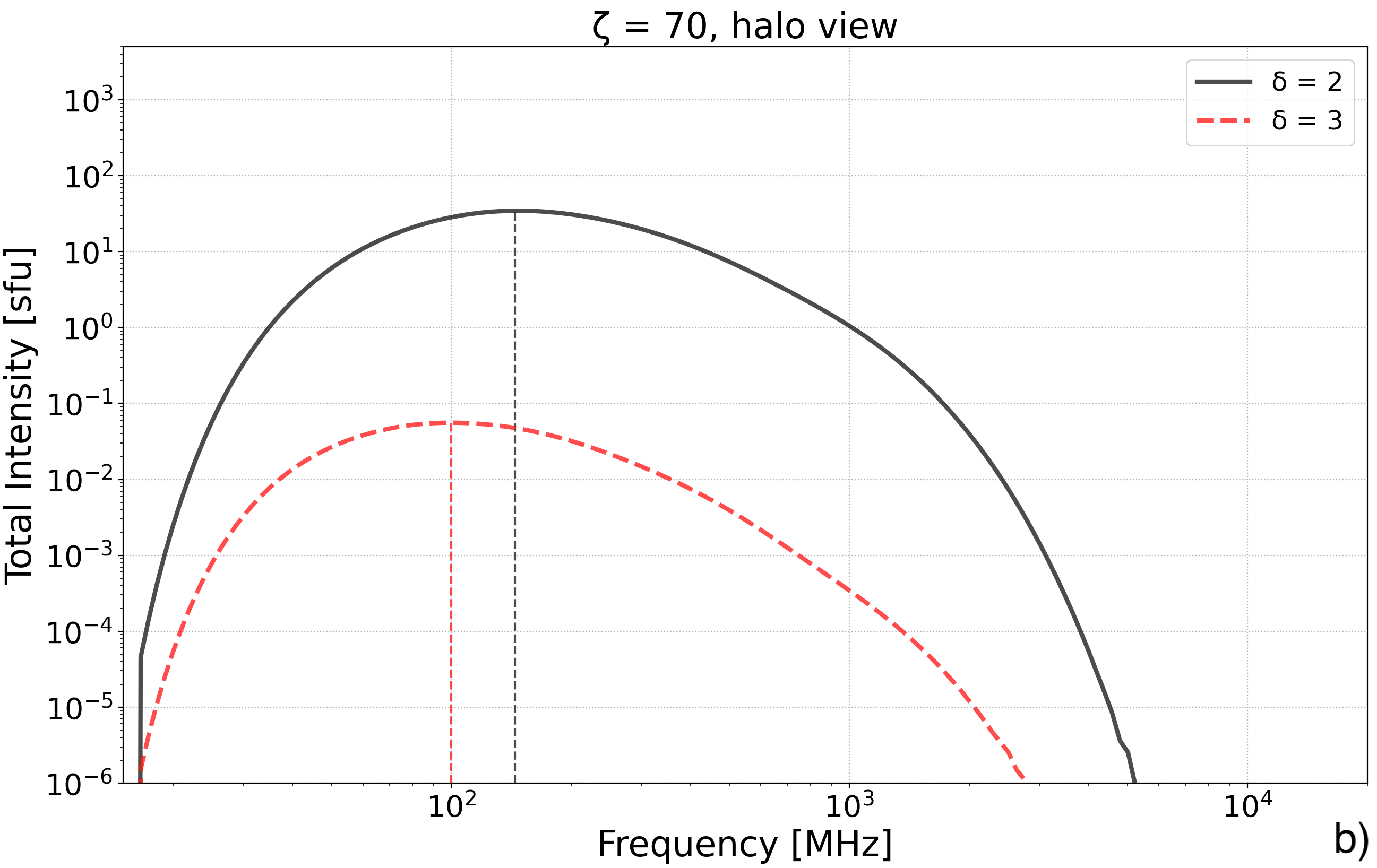} \\
        \includegraphics[width=0.45\textwidth]{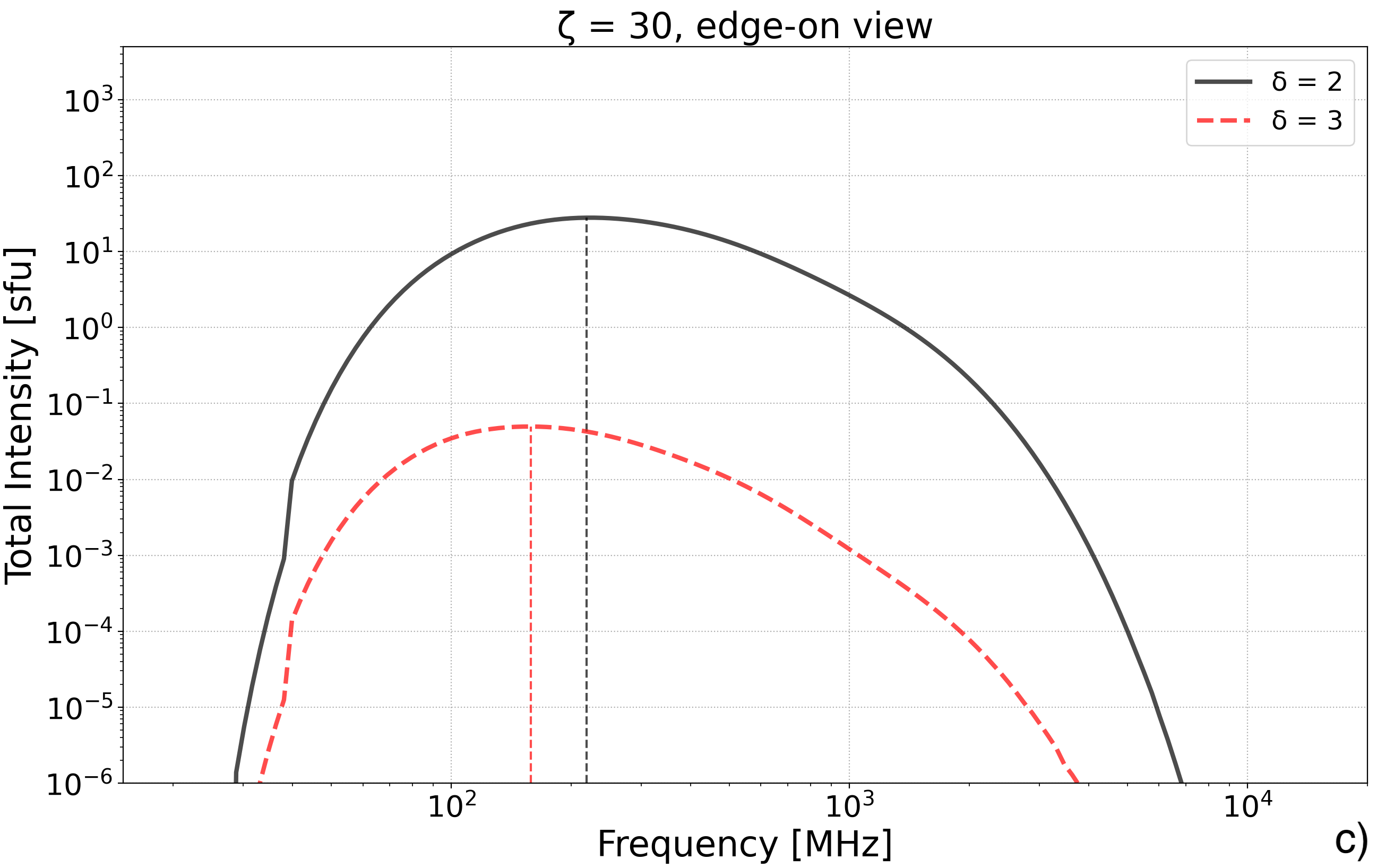} & 
        \includegraphics[width=0.45\textwidth]{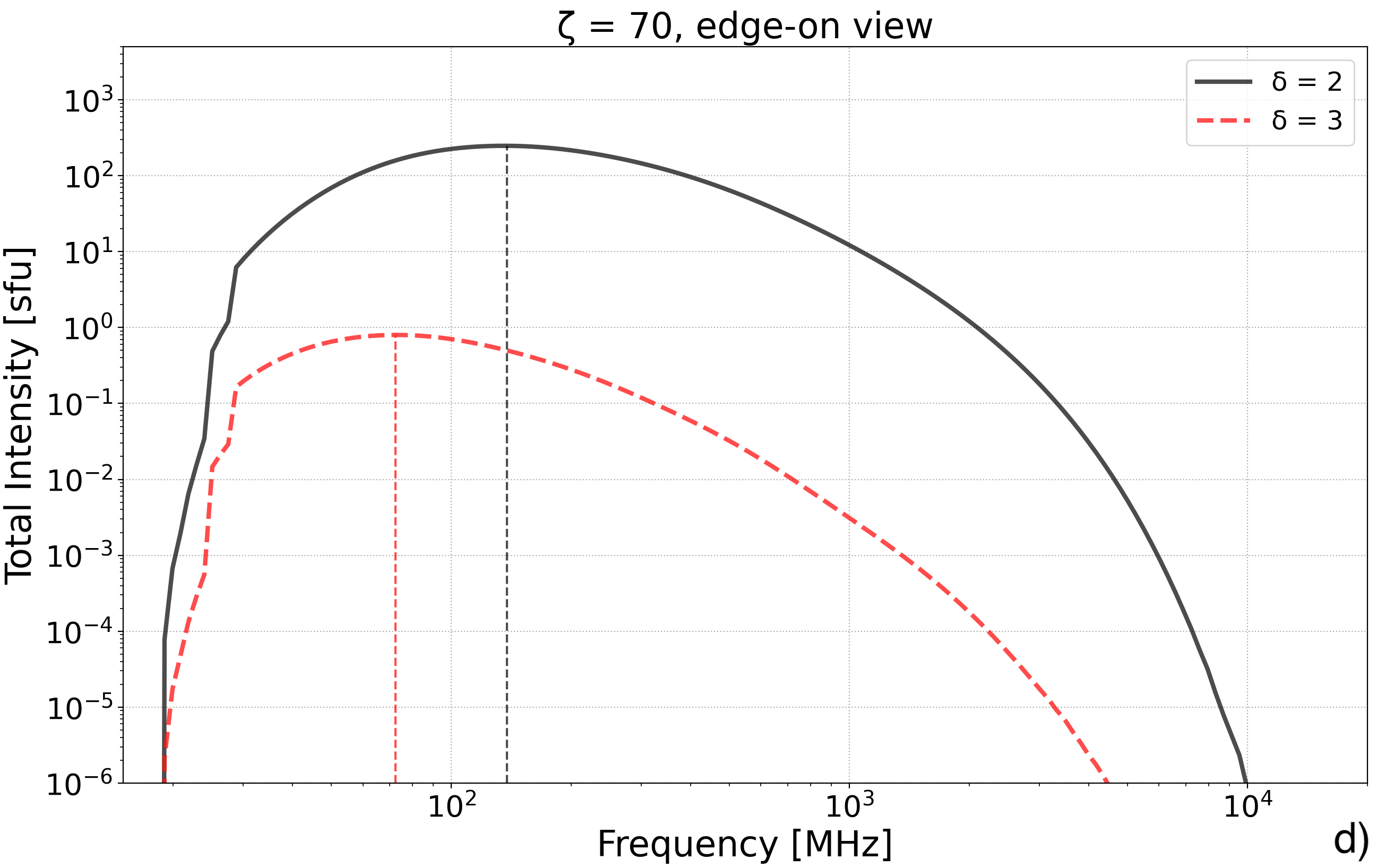} \\
        \includegraphics[width=0.45\textwidth]{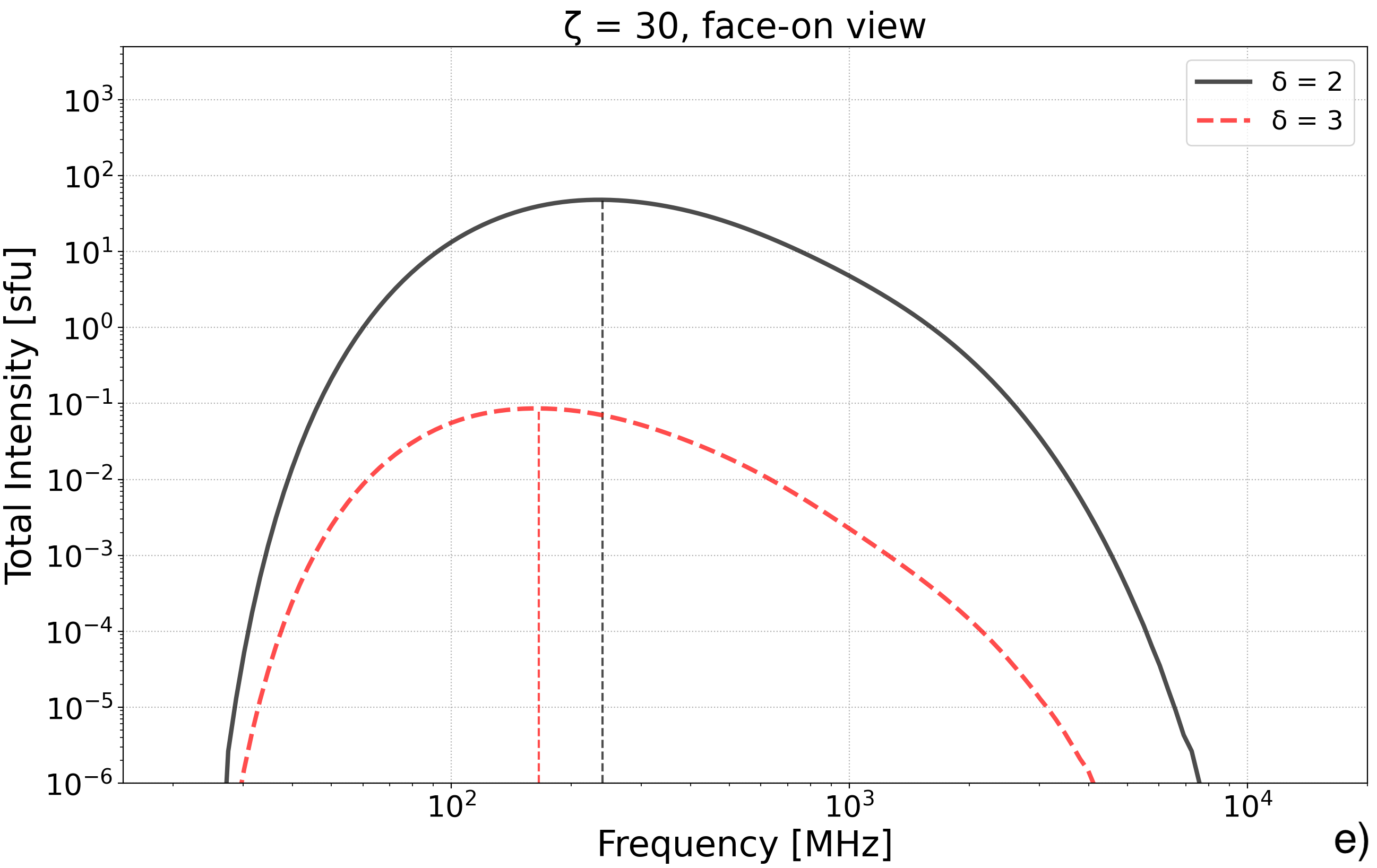} & 
        \includegraphics[width=0.45\textwidth]{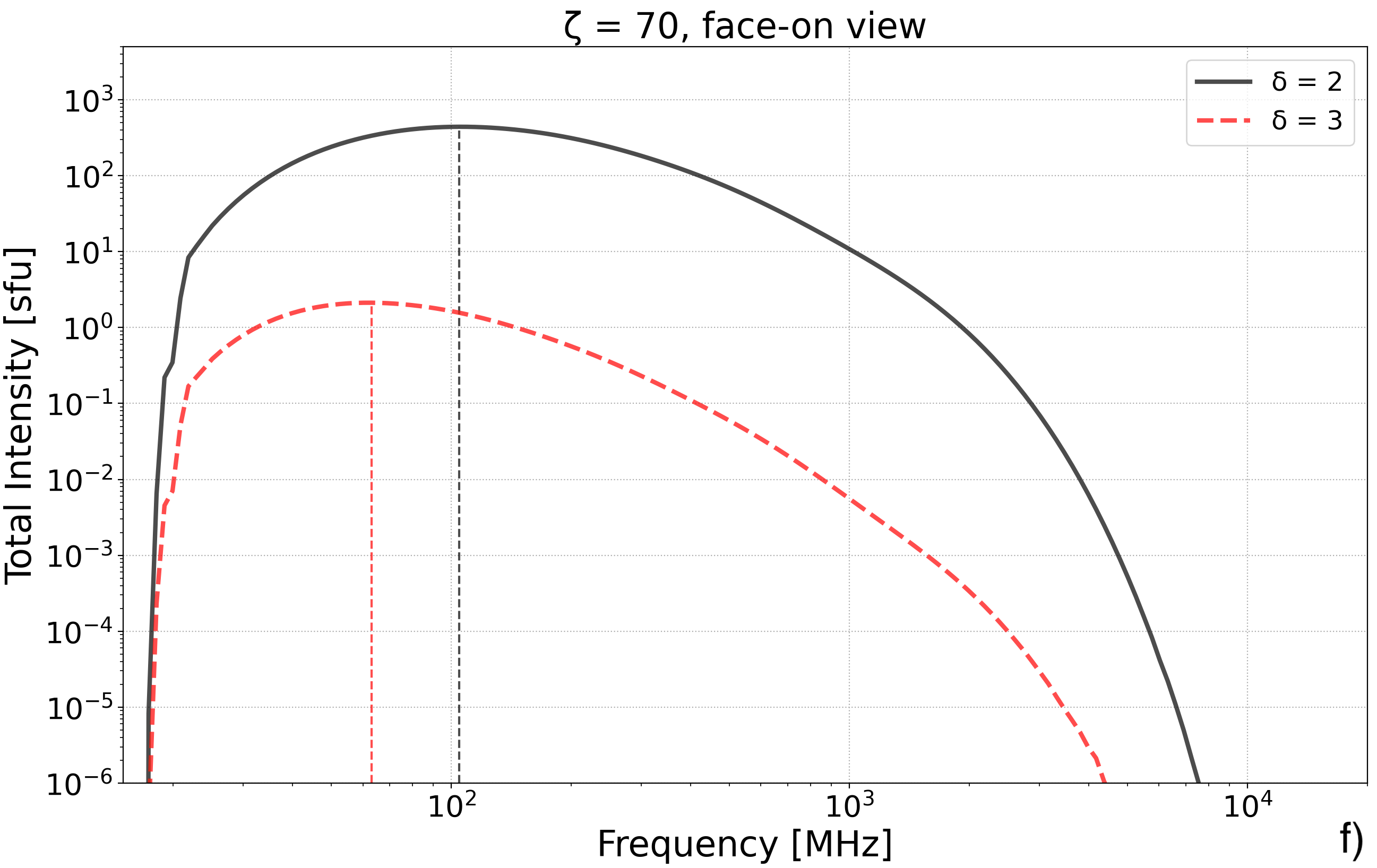} \\
    \end{tabular}
    \caption{Intensity curves as a function of frequency for a single time. The panels in the left column show the emission spectra in the case $\zeta = 30$ of the electrons injected with spectral indices $\delta = 2$ (black solid line) and $\delta = 3$ (red dashed line). In contrast, the right column depicts the equivalent results in the case $\zeta = 70$. Vertical lines mark the roll-over (or peak) frequencies. All spectra were recorded approximately 32 minutes into the simulation, which is about 3 minutes after the injection of the electrons.}
    \label{fig:curves}
\end{figure*}

A clear positive correlation is observed between stronger magnetic fields and higher GS emission intensities. The highest emission levels occur in the $\zeta = 70$ CME simulation for the face-on and edge-on views, which coincide with the strongest weighted magnetic fields in Fig.~\ref{fig:B_field}. Additionally, the halo view observer in the $\zeta = 70$ case registers stronger magnetic fields than the halo and edge-on view observers in the $\zeta = 30$ simulation, consistent with the observed peak GS intensities (see also Fig.~\ref{fig:curves}). 

However, the magnetic fields observed in the $\zeta = 70$ halo view are initially weaker than those in the $\zeta = 30$ face-on view, which is consistent with the detection of slightly lower GS intensities; however, this trend reverses at later times. Notably, the rising trend of the $\zeta = 70$ halo view curve does not necessarily indicate an increase in the CME's field strength but may result from CME expansion. When comparing field strengths along individual halo view LOS, we observe that increasing values occur along LOS farther from $90^\circ$ colatitude and $180^\circ$ longitude (i.e., the central launch coordinates of the CME) and into which the CME expands. In contrast, LOS already intersecting the CME and closer to these coordinates show a declining field strength over time. As a result, the final curve in Fig.~\ref{fig:B_field} reflects the growing contribution of oblique LOS contributing to the overall averaging.

\begin{figure*}
    \centering
    \begin{tabular}{cc} 
        \includegraphics[width=0.49\textwidth]{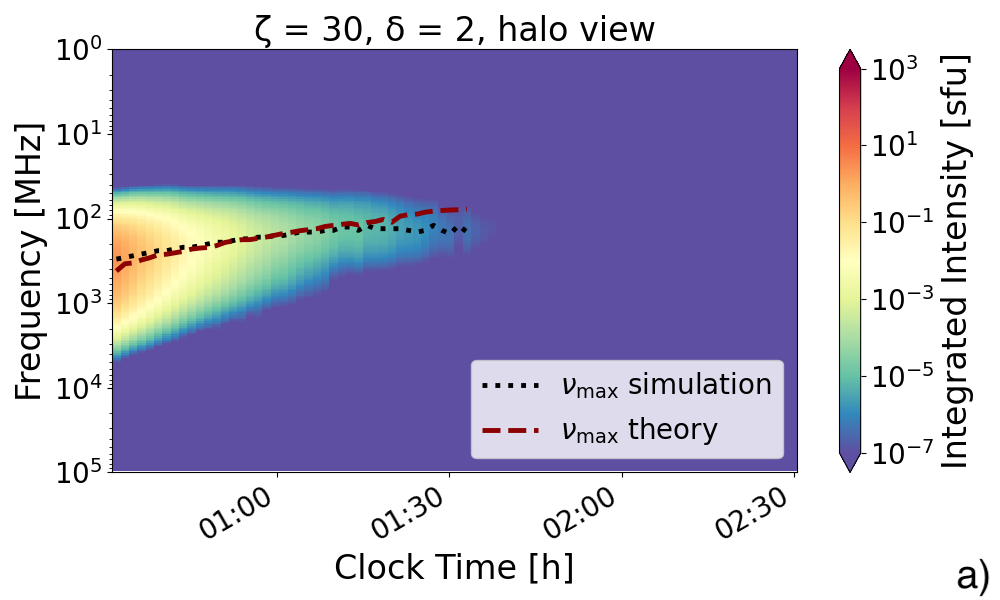} & 
        \includegraphics[width=0.49\textwidth]{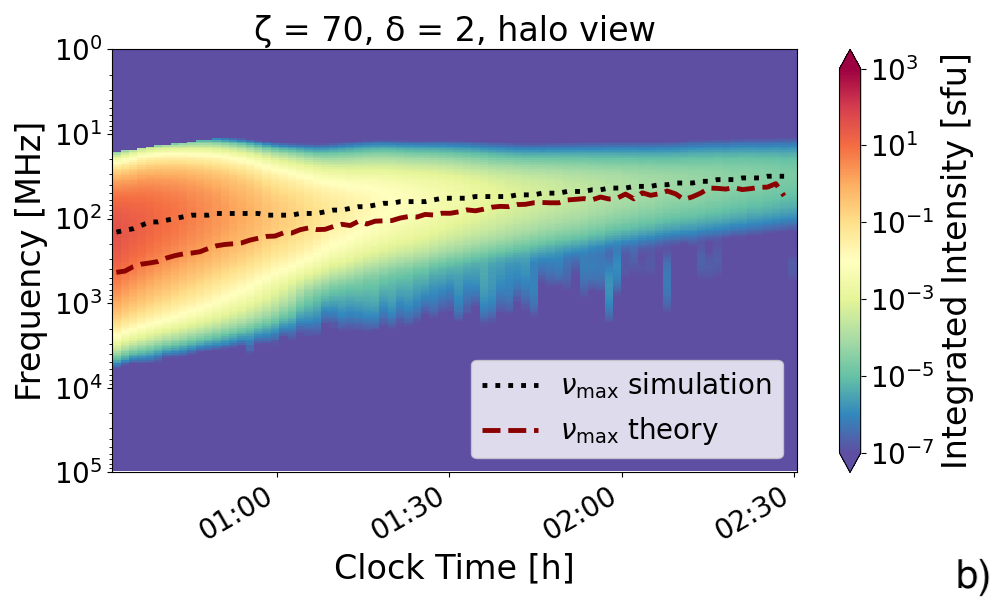} \\
    \end{tabular}
    \caption{Radio spectra with the peak intensity drifts from the simulation results and those calculated from theory. Exemplary, the drift of the peak intensity from higher to lower frequencies with time is shown for the case $\zeta = 30$, $\delta = 2$ (left panel), and $\zeta = 70$, $\delta = 2$ (right panel), both seen from the halo view. The black dotted curve shows the peak intensities from the simulation results, while the dark-red dashed curve shows the theoretical peak intensities calculated with Eq.~\ref{eq:nu_peak}.}
    \label{fig:validation_spectra}
\end{figure*}

As described earlier, the simulations both in Fig.~\ref{fig:radio_spectra_zeta_30} and \ref{fig:radio_spectra_zeta_70} indicate a shift of the peak intensities from higher to lower frequencies with time. To illustrate this frequency drift, we present two examples in Fig.~\ref{fig:validation_spectra}, where the left panel shows the case $\zeta = 30$ (same as the top left panel in Fig.~\ref{fig:radio_spectra_zeta_30}), and the right panel shows the case $\zeta = 70$ (same as the top left panel in Fig.~\ref{fig:radio_spectra_zeta_70}). Both panels correspond to the halo view and are based on $\delta = 2$. The black dotted curve indicates the peak intensities in the observed radio spectrum at each time step. To compare the results with theoretical expectations and to serve as a form of model validation, we additionally calculated the characteristic synchrotron peak frequencies following \cite{Ginzburg-1979} and \cite{Longair-1992} as 
\begin{align}
    \nu_\mathrm{peak}(t) = \frac{3}{4\,\pi} \frac{e\,B(t)}{m_\mathrm{e}\,c}\,\gamma^2 = \frac{3}{4\,\pi} \, \nu_\mathrm{g}(t)\,\gamma^2\,, \label{eq:nu_peak}
\end{align}
which are included as dark-red dashed curves in Fig~\ref{fig:validation_spectra}. For the magnetic field in Eq.~\eqref{eq:nu_peak}, we calculated a linearly weighted mean of $B$ along each line of sight, using normalised particle intensities as weights. The simulation and theoretical peak intensities are in reasonable agreement, showing a similar trend. We note that Eq.~\eqref{eq:nu_peak} is formulated for a single (constant) Lorentz factor. Since the highest-energy electrons contribute most to the GS emission in type~IV radio bursts, we computed the Lorentz factor based on an electron kinetic energy of 9~MeV. However, choosing lower energies can significantly alter the theoretical curve, as the Lorentz factor in Eq.~\eqref{eq:nu_peak} is a constant scaling factor. For instance, using a kinetic energy of 6~MeV instead of 9~MeV would reduce the theoretical frequencies to approximately $47\%$ of their original values.  

\begin{table}[ht]
\centering
\caption{Peak intensity ratios across observer positions, spectral indices, and CME cases.}
\begin{tabular}{llll}
\hline
\multicolumn{4}{l}{(1) Intensity ratios across views} \\
$\zeta$ & $\delta$ & $I_{\mathrm{edge-on}}/I_{\mathrm{halo}}$ & $I_{\mathrm{face-on}}/I_{\mathrm{halo}}$ \\
\hline
30 & 2 & 13.1 & 22.5 \\
30 & 3 & 25.5 & 44.1 \\
70 & 2 & 7.2 & 12.8 \\
70 & 3 & 14.3 & 38.1 \\
\hline
\multicolumn{4}{l}{(2) Intensity ratios across spectral indices} \\
$\zeta$ & View & $I_{\delta = 2}/I_{\delta = 3}$ \\
\hline
30 & halo & 1100 \\
              & edge-on & 564 \\
              & face-on & 560 \\
70 & halo & 620 \\
              & edge-on & 310 \\
              & face-on & 208 \\
\hline
\multicolumn{4}{l}{(3) Intensity ratios across CME cases} \\
$\delta$ & View & $I_{\zeta = 70}/I_{\zeta = 30}$ \\
\hline
2 & halo & 16.2 \\
              & edge-on & 8.9 \\
              & face-on & 9.2 \\
3 & halo & 28.6 \\
              & edge-on & 16.1 \\
              & face-on & 24.7 \\
\hline
\end{tabular}
\label{tab:intensity_ratios}
\end{table}

Finally, we illustrate in Fig.~\ref{fig:curves} single-line profiles of the observed intensities as a function of frequency, approximately 32 minutes into the simulation and about 3 minutes after the electron injection. The left column shows $\zeta = 30$ results, while the right column corresponds to $\zeta = 70$, with rows ordered according to the three spacecraft positions. Each panel contains results based on the two spectral indices $\delta = 2$ (black solid curve) and $\delta = 3$ (red dashed curve). As noted in the discussion of the histograms above, the smaller spectral index (i.e., a flatter slope) consistently produced higher intensities. 

Furthermore, in all cases, the roll-over (i.e., the transition from the optically thick, self-absorbed regime to the optically thin, emission-dominated regime), indicated by the vertical lines, occurs at higher frequencies for the $\delta = 2$ electron spectrum compared to the $\delta = 3$ spectrum. The roll-overs are found between 165~MHz and 300~MHz in the $\zeta = 30$ simulation, and between 65~MHz and 145~MHz in the $\zeta = 70$ simulation, with those for the smaller spectral index being shifted to higher frequencies by approximately 40~MHz to 70~MHz.

Comparing the spectra across the spacecraft positions for each CME case also reveals a shift of the roll-over frequency to lower values for vantage points where stronger GS emission was observed. When comparing the results across the two CME cases, we find that for $\zeta = 70$ (a stronger magnetic field), the roll-over occurs at lower frequencies (approximately 65~MHz to 145~MHz) compared to the corresponding $\zeta = 30$ case (approximately 165~MHz to 300~MHz). 

At first glance, this contradicts Eq.~\eqref{eq:nu_peak}, since the frequency at the intensity peak scales with the electron gyrofrequency and thus with the magnetic field strength. However, the spectra in Fig.~\ref{fig:curves} show that the slopes of the intensity curves in the $\zeta = 70$ case are steeper than in the $\zeta = 30$ case, suggesting stronger self-absorption, which shifts the roll-over to lower frequencies, reducing the transition to the optically thin regime.

Although the intensity curves displayed in Fig.~\ref{fig:curves} do not have a clear power-law form, they nonetheless suggest spectral indices $\chi$ that are steeper than those predicted by conventional analytical models of synchrotron and GS emission.
In the optically thin regime, the typical relation for synchrotron emission is \(\chi = (\delta - 1)/2\) \citep{Ginzburg-Syrovatskii-1964}, while a commonly used approximation for GS emission is \(\chi = 0.9\,\delta - 1.22\) \citep{Dulk-Marsh-1982}. As expected, the GS slope is steeper than that of pure synchrotron emission, likely due to propagation effects and the influence of the ambient medium. At the same time, these idealised expressions do not fully capture the complexities of real systems. Observational data frequently reveal significantly steeper indices, and our particle transport simulations are consistent with this pattern, showing early deviations from the initial spectral index of the injected electron distributions towards steeper slopes. This steepening may, in part, be explained by energy-dependent transport effects. Lower-energy electrons (e.g., $\sim 100$~keV) have longer effective residence times in the flux rope due to their smaller velocities and reduced mirroring rates. In contrast, higher-energy electrons undergo more frequent magnetic mirroring between footpoints, increasing their probability of entering the loss cone and precipitating. Thus, their reduced contribution to the observed emission can lead to a steeper spectral slope. Additional discrepancies between analytical predictions and our numerical results may arise from uncertainties in the ambient plasma properties and simplifications inherent in analytical models.

While we considered only (incoherent) GS emission in this work, the presence of coherent emission mechanisms, often proposed for type~IV bursts, cannot be ruled out. Plasma emission, in particular, has been suggested for both type~IVs \citep{Weiss-1963, Benz-etal-1976} and type~IVm bursts \citep{Gary-etal-1985, Morosan-etal-2019}. In addition, maser emission driven by a maser instability in nonthermal electron populations with strong perpendicular anisotropy, typically characterised by a loss-cone distribution, has frequently been proposed as a contributing mechanism \citep{Winglee-Dulk-1986, Aschwanden-Benz-1988, Treumann-etal-2011, Morosan-etal-2016}. Under suitable conditions, such instabilities can amplify electromagnetic waves near the electron gyrofrequency (e.g., \citealt{Tsytovich-1970, Papadopoulos-Freund-1979, Galeev-Krasnoselskikh-1979}). Whether such plasma instabilities actually develop in coronal conditions similar to those modelled here, and whether the resulting coherent emission is present in observed type~IV radio spectra, remains an open and debated question (e.g., \citealt{Morosan-etal-2016,Carley-etal-2017}).

\section{Summary and conclusions}\label{sec:summary}

This study presented a novel coupling of three models to generate synthetic radio spectra and investigate the influence of different injection spectra and initial CME parameters on the observed emission spectra. Using the 3D MHD coronal model COCONUT, we constructed a coronal configuration based on an HMI magnetogram. We then performed two simulations, each modelling a CME as an MFR with different initial properties in terms of magnetic field strength and speed, encapsulated in the dimensionless parameter $\zeta$, here with values of  30 and 70. Next, we used the SEP model PARADISE to simulate energetic electrons trapped within the erupting MFR. Finally, using COCONUT's plasma parameters and the electron energy distributions obtained from PARADISE, we applied the UFGSCs to compute GS emission from the trapped electron populations as observed from three different vantage points ("halo view", "edge-on view", "face-on view") in the solar corona.

Comparing the radio spectra for each CME across different spectral indices, we found that the flatter spectrum with $\delta = 2$ consistently produced stronger GS radiation and longer-lasting radio bursts than the steeper spectrum with $\delta = 3$. A comparison between the two CMEs revealed that the electrons in the $\zeta = 70$ CME, characterised by a stronger initial magnetic field strength (nearly twice that of the $\zeta = 30$ CME) and a higher initial speed, generated much more intense GS emission. Finally, benchmarking the three observer positions against each other demonstrated that the strongest GS emission was detected when the spacecraft was above the CME, followed by the perspective viewing both CME flanks from edge-on. In contrast, the weakest GS emission was detected when the observer faced the front of the approaching CME (halo view). 

It should be noted that modifying the employed energy range or further steepening the injection spectrum would significantly affect the resulting GS emission, including frequency range, intensity, and duration. Since an electron's pitch angle directly influences its radiated power, the initial pitch-angle distribution of the injected population can also impact the morphology and quantitative properties of the resulting type~IV spectra. While we adopted an isotropic distribution in this study, test simulations with highly beamed electron populations revealed a range of pitch-angle distributions at early times in the type~IV event, depending on the energy channel and location. For instance, we found single electron beams (both field and anti-field aligned), counter-beam populations, and populations transitioning towards isotropy. The computed spectra were morphologically similar, but showed slightly reduced peak intensities (by $\sim 20$~\%). This is consistent with theoretical expectations for smaller pitch angles. Furthermore, the reduced GS emission may also stem from a significant fraction of particles entering the magnetic bottle's loss cone, leading to precipitation into the inner simulation boundary and their removal from the simulation.

A detailed exploration of these dependencies goes beyond the scope of the present study. In future work, we aim to conduct MHD and transport simulations with significantly higher temporal resolution (e.g., 10~s sampling times), focusing on the early evolution (first $\sim$30--60~min) of type~IV events to investigate the role of pitch-angle and other energy distributions (e.g., initially monoenergetic or kappa-distributed, or using different energy ranges) more systematically.

The radio spectra obtained from our simulations suggest that type~IV radio bursts contain GS radiation from electrons trapped within the strong magnetic field lines of CMEs. Furthermore, our results indicate that the GS emission primarily originates from the CME flanks, consistent with imaging observations and previous studies (see, e.g., \citealt{Gopalswamy-Kundu-1990,Bastian-etal-2001,Bain-etal-2014, Carley-etal-2020}). Additionally, our simulations demonstrate that the orientation of the flux rope relative to the observer's LOS significantly influences the detected emission (e.g., \citealt{Nindos-2020}). This finding highlights the importance of multi-messenger observations, as the evolution of the radio source cannot be fully understood without complementary observational constraints.

All computed emission spectra exhibit a similar structure, consisting of a high-intensity core surrounded by regions of weaker emission, with the peak intensity gradually drifting to lower frequencies over time. These spectral features are physical within the GS framework implemented in the UFGSCs and arise entirely from incoherent GS emission. The spatial and spectral variations, such as LOS-dependent intensity patterns and secondary emission lanes, can be consistently understood using the Li\'{e}nard formula \citep{Schwinger-1949}. GS emission is particularly enhanced in MFR flank regions near the footpoints, where the local pitch-angle distribution and magnetic field geometry favour spiky, burst-like signatures. These regions naturally support GS radiation, as similar behaviour is expected whenever electrons are injected into an MFR, though the absolute brightness depends on the flux rope structure itself.

While the present work did not attempt a case study, several aspects support the physical realism of our simulation results. All three coupled models are physics-based, providing physical credibility beyond empirical parameter fitting. Moreover, the morphological features of the obtained spectra match type~IV observations. Finally, our results quantitatively fall within observationally established ranges in terms of frequency intervals, spectral flux densities, frequency drift rates (supported both by literature values and theoretical expectations), and event durations.

In conclusion, our simulation results yield several key insights regarding GS emission in type~IV radio bursts:
\begin{enumerate}
    \item A synchrotron / GS background is likely present in most type~IV radio bursts. The brightness of this background is determined by the high-energy electron distribution, the magnetic field strength, the topology, and other properties of the MFR confining the electrons, as well as its orientation relative to the observer.
    \item GS emission is strongest in the CME flanks near the footpoints rather than the CME apex, due to stronger magnetic field strengths and more favourable pitch-angle distributions, supporting earlier observations of enhanced GS emission from CME flanks.
    \item Both spectral and spatial drift are natural consequences of CME expansion and evolving field conditions. In our simulations, the GS emission source (i.e., energetic electrons) moves outwards with the expanding MFR, while the associated spectral drift depends on the magnetic field evolution and energetic electron distribution. This may help explain why some type~IV bursts are observed as moving, while others seem stationary in radio imaging.
\end{enumerate}
These findings emphasise the importance of considering GS emission as a fundamental mechanism in type~IV radio bursts, and illustrate how CME properties and observer geometry shape the observed radio spectra. Additionally, our results suggest that type~IVm spectra can exhibit drift rates below thresholds proposed in previous studies (e.g., \citealt{Kumari-etal-2021}), implying that some bursts classified as IVs might still be associated with erupting MFRs. While our analysis is solely based on incoherent GS radiation, it does not necessarily exclude the presence of other emission mechanisms, which may contribute or modify the observed spectral features. In particular, the observed spectral flux densities may result from a superposition of multiple mechanisms, including coherent processes, which may yield stronger emission than the GS component alone in our simulations. Ultimately, this study helps clarify the role of GS emission in shaping radio spectra and provides a framework for identifying regions where additional coherent emission from the maser instability might occur, which is beyond the scope of this work.

The present study provides a foundation for future research on CME kinematics and GS emission to infer electron properties and magnetic field characteristics within CMEs, with the potential for direct validation using PSP's radio and particle measurements. Future work will also explore magnetoionic properties of GS radiation, focusing on polarisation to refine estimates of magnetic field strength and topology.

\begin{acknowledgements}
We thank the anonymous referee for their valuable comments, which helped improve the clarity and rigour of the manuscript.
Computational resources and services used in this work were provided by the VSC (Flemish Supercomputer Center), funded by the Research Foundation - Flanders (FWO) and the Flemish Government  – department EWI.
E.H.\ is grateful to the Space Weather Awareness Training 
Network (SWATNet) funded by the European Union's Horizon 2020 
research and innovation program under the Marie 
Skłodowska-Curie grant agreement No. 955620.
N.W.\ acknowledges funding from the
KU Leuven project 3E241013. 
I.C.J. is grateful for support by the Research Council of Finland (SHOCKSEE, grant No.~346902, and X-Scale, grant No.~371569), and the European Union’s (E.U's) Horizon 2020 research and innovation program under grant agreement No.\ 101134999 (SOLER). A.V. is supported by NASA grant 80NSSC21K1860. 
S.P.\ is funded by the European Union. Views and opinions expressed are, however, those of the author(s) only and do not necessarily reflect those of the European Union or ERCEA. Neither the European Union nor the granting authority can be held responsible for them. This project (Open SESAME) has received funding under the Horizon Europe programme (ERC-AdG agreement No 101141362). These results were also obtained in the framework of the projects C16/24/010 C1 project Internal Funds KU Leuven), G0B5823N and G002523N (WEAVE) (FWO-Vlaanderen), and 4000145223 (SIDC Data Exploitation (SIDEX2), ESA Prodex).
\end{acknowledgements}

\bibliographystyle{aa}
\bibliography{gs_paper}

\begin{appendix} 

\section{Technical details of the numerical models}\label{app: technical details}
\subsection{COCONUT}\label{app: COCONUT}

The coronal MHD model COCONUT solves the 3D ideal MHD equations to generate coronal configurations from 1 to 21.5\,$R_\odot$ \citep{Perri-etal-2022}. To model CMEs, we use the analytical modified Titov--D\'{e}moulin flux rope model \citep{Titov-Demoulin-1999, Titov-etal-2014, Linan-etal-2023}, in the following as MFR addressed. An eruptive MFR is achieved by setting the ring current intensity $I$ in the MFR to a value larger than the Shafranov intensity $I_\mathrm{S}$ via
\begin{align}
    I = \zeta\,I_\mathrm{S}  \approx \zeta \, \frac{R\,B_\mathrm{t}}{\ln{\frac{8\,R}{a}} - \frac{3}{2} + \frac{I_\mathrm{i}}{2}}\,, \label{eq:shafranov}
\end{align}
which causes the magnetic pressure inside the MFR to exceed that of the ambient coronal plasma. In Eq.~\eqref{eq:shafranov}, $\zeta$ is a dimensionless parameter, $R$ and $a$ describe the major and minor radius of the flux rope, respectively, $B_\mathrm{t}$ is the toroidal magnetic field component, and $I_\mathrm{i}$ is a dimensionless variable that characterises the radial distribution of the poloidal magnetic field inside the MFR (for details, see \citealt{Titov-etal-2014,Linan-etal-2023}).

The initial placement of the CME footpoints (i.e., the locations where the MFR is magnetically anchored in the lower corona or photosphere) is around colatitude $\theta = 90^\circ$ and Carrington longitude $\phi = 180^\circ$, with the centre of the toroidal axis located at a height $d = 0.15\,R_\odot$ above the solar surface. The minor radius is set to $a = 0.1~R_\odot$, while the major radius to $R = 0.3~R_\odot$. The unstructured COCONUT grid consists of about 2 million prism-shaped cells arranged in concentric shells and increasing in size with radial distance (see \citealt{Brchnelova-etal-2022} for details). The output cadence of the MHD snapshots is about 87 seconds, and we use the first 2.5 hours of the COCONUT simulations for the subsequent particle transport simulations. We further note that the actual computational domain of COCONUT in our simulations extends to 25~$R_\odot$, but we set the outer boundary in the subsequent particle transport simulations to $21.5$~$R_\odot$ to avoid any numerical outer boundary effects, following the recommendations by \cite{Brchnelova-etal-2022}.

\subsection{PARADISE}\label{app: PARADISE}

The particle transport code PARADISE solves the FTE in the present work in the form
\begin{align}
\frac{\partial f}{\partial t} 
+ \frac{\mathrm{d}\mathbf{x}}{\mathrm{d}t} \cdot \nabla f 
+ \frac{\mathrm{d}\mu}{\mathrm{\mathrm{d}}t} \frac{\partial f}{\partial \mu} 
+ \frac{\mathrm{d}p}{\mathrm{d}t} \frac{\partial f}{\partial p} 
= \frac{\partial}{\partial \mu} \left( D_{\mu\mu} \frac{\partial f}{\partial \mu} \right)
\label{eq:FTE}
\end{align}
to derive spatio-temporal particle intensity distributions. In Eq.~\ref{eq:FTE}, $\mathbf{x}$ is the spatial vector, and $D_{\mu\mu}$ the pitch-angle diffusion coefficient. Detailed expressions for the single terms in the FTE~\ref{eq:FTE} describing effects such as adiabatic cooling, focussing and mirroring can be found, for instance, in \cite{Wijsen-2020} and \cite{le-Roux-Webb-2009}, while information about $D_{\mu\mu}$ can be found in \cite{Agueda-Vainio-2013}.

To ensure robust statistics in the particle transport simulations, a total of 10.8 million electrons are injected in each simulation. While the pitch-angle diffusion coefficient is included in the FTE, for simplicity, we exclude cross-field (perpendicular) diffusion, and the guiding-centre drift velocity is set to zero, as it is expected to be negligible for electrons. The parallel mean free path length is fixed to $\lambda_\parallel = 0.1$~au. Further details about the coupling between COCONUT and PARADISE can be found in \cite{Husidic-etal-2024}.

At each node along a line of sight and for each time step, the particle distribution function $f$, obtained from PARADISE, is provided to the UFGSCs as a 2D array of values $f_{i,j} = f(E_i, \mu_j)$ defined in energy-pitch-angle space. Here, $E$ denotes the kinetic energy, and $f$ has the units cm$^{-3}$\,MeV$^{-1}$. We add a background electron component to the PARADISE output to avoid exact zeros resulting from the logarithmic energy grid. This background is based on the relativistic regularised kappa distribution (rRKD; \citealt{Han_Tanh-etal-2022}), reformulated in terms of relativistic kinetic energy $E = \sqrt{(p\,c)^2 + (m_0\,c^2)^2} - m_0\,c^2$ with electron rest mass $m_0$, and provided in units cm$^{-3}\,$MeV$^{-1}$. The local number density and temperature at a given time and position in the rRKD are extracted from the MHD simulation. 

\begin{figure}
    \includegraphics[width=0.5\textwidth]{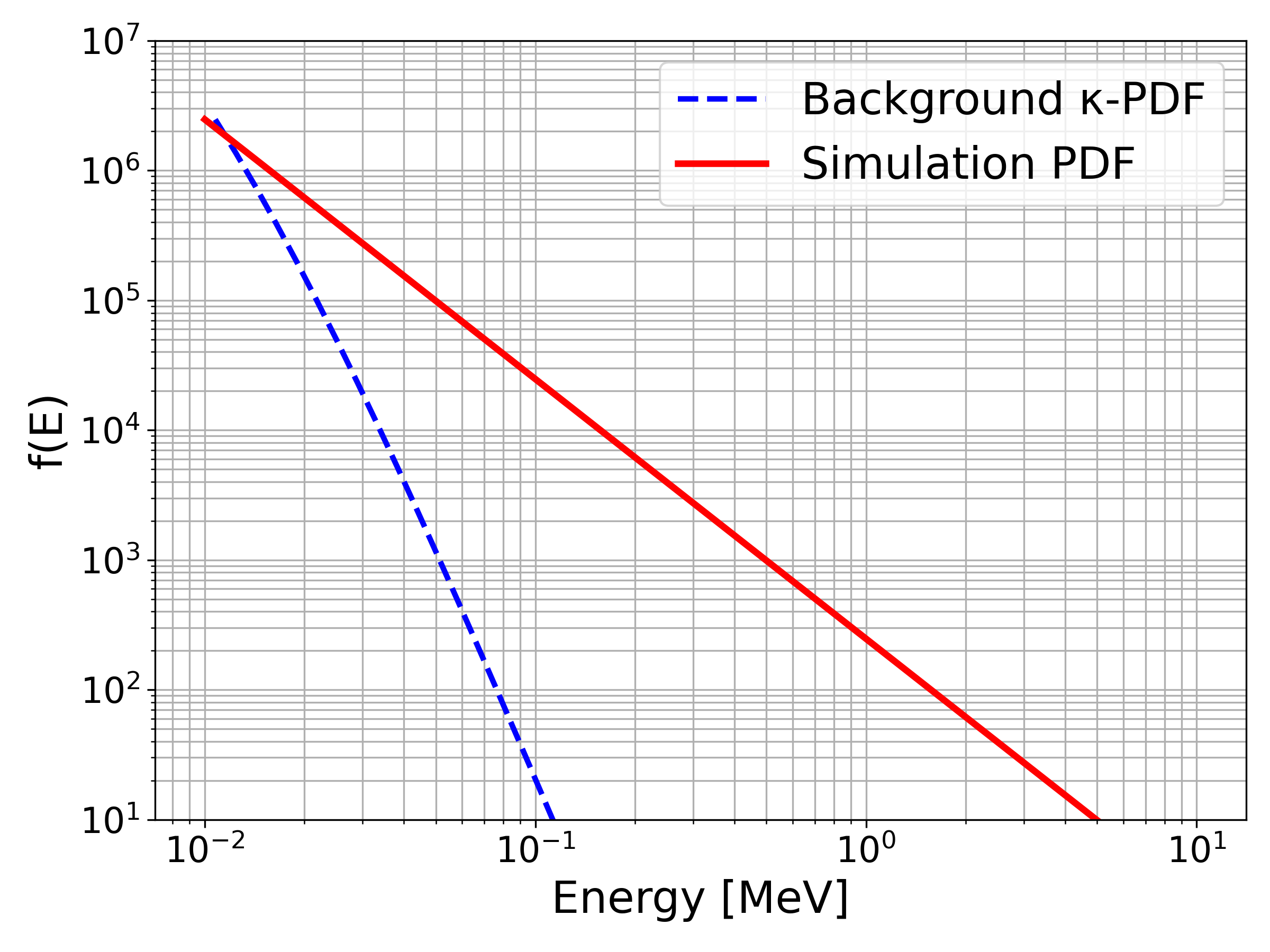}
   \caption{Comparing the distribution function tails of the background and the injected electron distribution. The particle distribution function (PDF) is plotted as a red solid graph, while the background rRKD with $\kappa = 8$ and $\xi = 0.001$ is plotted as a blue dashed graph.}
    \label{fig:PDFs}%
\end{figure}

To ensure that the PARADISE electron distribution has physically meaningful units, we obtain the scaling factor by matching the injected particle distribution at 10~keV to the rRKD, parametrised by the local solar wind conditions obtained from the MHD simulation at the injection region and time. The influence of the background is minimised in all simulations by choosing $\kappa = 8$ and a cut-off parameter $\xi = 0.001$ for the rRKD (for details, see \citealt{Scherer-etal-2017,Han_Tanh-etal-2022}). Figure~\ref{fig:PDFs} shows the injected electron distribution (red solid curve) for $\delta = 2$ and the background rRKD (blue dashed curve), demonstrating the scaling. As illustrated, the background distribution aligns with the PARADISE simulation at the lowest energy of the injection spectrum and remains below it across the whole energy range. Since the PARADISE distribution steepens during the transport simulation, selecting a $\kappa$-value of 8 ensures the background stays below. This also helps to avoid excessively steep gradients in the total distribution, which could otherwise lead to numerical artefacts in the GS calculation and provide conditions resulting in occasional oversaturated pixels in the computed spectra, with intensities sometimes exceeding those of neighbouring pixels by several tens of orders of magnitude. 

Including the background inevitably introduces a small GS contribution from it. However, due to the nonlinearity of the emission and absorption coefficients, the background contribution cannot be directly subtracted. Nevertheless, simulations run with the background distribution alone show that subtracting its GS emission from that of the combined (PARADISE $+$ background) distribution results in a change of less than 0.01\% at the peak intensities.

\end{appendix}

\end{document}